\documentclass[12pt, prd, nofootinbib, longbibliography ]{revtex4-2}
\usepackage{amsmath}
\usepackage{amssymb}
\usepackage{graphics}
\usepackage{graphicx}
\usepackage[final]{hyperref} % adds hyper links inside the generated pdf file

\newcommand{\ex}[1]{\langle x^{#1} \rangle}
\newcommand{\er}[1]{\langle r^{#1} \rangle}

\newcommand{\RR}{\mathbb{R}}
\newcommand{\HH}{\mathcal{H}}
\newcommand{\Rp}{\mathbb{R}_{+}}
\newcommand{\OO}{\mathcal{O}}

\newcommand{\ket}[1]{|#1 \rangle}
\newcommand{\Ai}{\text{Ai}}
\newcommand{\vev}[1]{\langle {#1} \rangle}
\newcommand{\pex}[1]{\langle {#1}\rangle_\psi}
\newcommand{\intplus}{\int_0^\infty dx\ }

\begin{document}

\title{Anomalous Bootstrap on the half line}
\author{David Berenstein $^\dagger$}
\author{George Hulsey $^\ddagger$}
\affiliation{Dept. of Physics, University of California, Santa Barbara\\Santa Barbara, CA 93106}

% more complex case: 4 authors, 3 institutions, 2 footnotes
% \author[a,b,1]{F. Irst,\note{Corresponding author.}}
% \author[c]{S. Econd,}
% \author[a,2]{T. Hird\note{Also at Some University.}}
% \author[a,2]{and Fourth}

% The "\note" macro will give a warning: "Ignoring empty anchor..."
% you can safely ignore it.
% e-mail addresses: one for each author, in the same order as the authors
\email{$^\dagger$ dberens@physics.ucsb.edu}
\email{$^\ddagger$ hulsey@physics.ucsb.edu}

\begin{abstract}
   We study carefully the problem of the bootstrap on the half line. We show why one needs the full set of constraints derived from the Stieltjes theorem on the moment problem by reexamining previous results on the hydrogen atom.
    We also study the hydrogen atom at continuous angular momentum. We show that the constraints on the moment problem alone do not fix the boundary conditions in all cases and at least one of the positive
     matrices needs to be slightly enlarged to remove unphysical branches. We explain how to solve the more general problem of the bootstrap for Robin boundary conditions. The recursion relations that are usually used receive additional anomalous contributions.  These corrections are necessary to compute the moments of the measure. We apply these to the linear potential and we show how the bootstrap matches the analytical results, based on the Airy function, for this example.
\end{abstract}

\maketitle

\section{Introduction}

In its simplest form, the quantum mechanical bootstrap consists of two steps: given some Hamiltonian for a system, compute moment sequences associated to its eigenvectors, then check if those moment sequences are consistent with a general positivity constraint of a truncated matrix of infinite size  \cite{Han:2020bkb}. Basically, one computes recursively moments 
\begin{equation}
\vev{{\OO_n}}= \vev{x^n} \label{eq:moments}
\end{equation}
assuming that the state is in an eigenstate of the Hamiltonian with energy $E$.  This procedure uses commutation relations of the operators $x^n, x^n p$ with the Hamiltonian to produce moment sequences for polynomial potentials from the energy plus any additional parameters that are required for initializing the sequence. This collection of parameters is called the search space.
One then asks  if these recursively computed sequences are consistent with the existence of a normalizable eigenstate solution of the Schr\"odinger equation. Any  linear combination operator ${\OO } \sim \sum_n a_n {\OO}_n$ must satisfy a positivity constraint
\begin{equation}
\langle \OO^\dagger\OO\rangle \geq 0 .\label{eq:square}
\end{equation}
This constraint can be thought of as a unitarity constraint: that the Hilbert space norm of ${\OO}\ket\psi$ is positive. 
 This constraint can be violated for some finite sequence $a_n$ if $E$ is not in the spectrum of the Hamiltonian, but this statement is not automatically guaranteed. 
 One can think that this failure is due to  missing some additional information which the sequence \eqref{eq:moments} is not capturing on its own.

  This procedure has been analyzed in a number of examples \cite{Berenstein:2021dyf, Bhattacharya:2021btd,  Aikawa:2021eai, Tchoumakov:2021mnh, Berenstein:2021loy, Aikawa:2021qbl}, which include some of our previous work in the subject. When the procedure works, one seems to get close to the correct values of $E$ exponentially fast in the size of the computed sequence of the $a_n$.  
From here one can guess and check for the allowed values of certain state parameters, like the energy or the value of specific positional moments and determine valid solutions of the bootstrap equations up to some value $n_{max}\equiv K$, which we call the depth of the test. 

How can we check if a moment sequence is allowed? This is answered by a set of questions (and answers) from the mathematical literature---the so-called classical moment problems. Given some possibly infinite interval $I \subseteq \RR$, the moment problem  is formulated  as follows: given a real sequence $a_n$, does there exist a positive measure $d\mu$ supported on $I$ such that $a_n = \int_I x^n d\mu$?

The three classical moment problems are those of Hamburger, Stieltjes, and Hausdorff, corresponding to the intervals $\RR,\Rp\cong [0,\infty)$ and $[0,1]$ respectively \cite{akh}. These are the three topological types of one-dimensional intervals. In a previous paper of ours, we numerically bootstrapped the spectrum of the hydrogen model to show the efficacy of the bootstrap method. However, we were unable to correctly obtain the s-wave states ($\ell = 0$). 

The reason for this was that we checked an incomplete set of constraints. The radial sector of the hydrogen model is quantum mechanics on the half line $\Rp$. Hence, checking for valid measures requires using the theorem of Stieltjes: 
\vspace{0.25cm}\\
\textbf{Stieltjes, 1894.} Let $\{a_n\}$ be a sequence of real numbers. The $a_n$ correspond to the moments of a normalizable measure $\mu$ on $\Rp$, i.e. $a_n = \int_0^\infty r^n\ d\mu$, if and only if the  two matrices with elements $M_{ij} = a_{i+j},\ \Tilde{M}_{ij} = a_{i+j+1}, 0 \leq i,j \leq K-1 $ are positive semi-definite for all $K$.
\vspace{0.25cm}\\
In this notation, we need $d\mu\geq 0$, so $\mu$ is a non-decreasing function. The measure $\mu$ is not necessarily unique; it is provided the $a_n$ don't grow too quickly. These growth conditions are usually satisfied by proper bound states in quantum mechanics.

The notable difference between this result and the result for the moment problem on $\RR$ is the positivity condition on the second matrix
$\Tilde{M}$ \footnote{Recall that the moment problem on $\RR$, the Hamburger problem, requires only positivity of the matrix with elements $M_{ij} = a_{i+j},\ 0 \leq i,j\leq K-1$ at all depths $K$.}. The necessity of this requirement follows from positivity of the norm: consider an operator $\OO = \sqrt{r}\sum c_nr^n$. Such an operator is well defined when the position operator is positive $r > 0$. Then, in any state, positivity of the expectation value $\langle \OO^\dagger \OO \rangle \geq 0$ implies the condition, for $\forall c_n$: 
\begin{equation*}
    \langle \OO^\dagger \OO \rangle = \sum_{n,m}c^*_n \langle r^{n+m+1}\rangle c_m \geq 0
\end{equation*}
This is equivalent to positive (semi)definiteness $\Tilde{M} \succeq 0$. Proving sufficiency of this condition is more difficult and is related to extensions of positive, symmetric operators \cite{reed2}.

The positivity condition on the second matrix introduces new constraints on the moments, leading to improved convergence of the numerical algorithm. While this generally improves the performance of the algorithm for problems on the half line, there remain aspects of the bootstrap for half-line problems which are not obviously addressed by the bootstrap problem as defined so far. Essentially, we need to understand the role and determination of boundary conditions. After we revisit our earlier work on the hydrogen model, where we show that this theorem addresses the shortcomings of our previous work, we introduce the Airy model. The results from the Airy bootstrap are intriguing as they betray some implicit assumptions about boundary conditions in the most na\"\i ve way of  computing the recursion relations for the $a_n$. This naturally leads to more technical discussion of the data that we supply the bootstrap and allows us to generate the terms required to specify boundary conditions. The main new understanding is that the recursion relations that are used to iteratively compute the moments from some initial data of the moments receive additional anomalous contributions. These anomalous terms arise from a failure of some boundary terms to vanish in mathematical manipulations that require integrating by parts. These same terms vanish naturally in the problem over $\RR$, because the measure decays sufficiently fast at infinity. 
The proper theory of why this happens over $\Rp$ has to do with domains of dependence of operators (understanding correctly the space of  functions on which the operators act). In this case we solve the problem of how to determine the recursion equations when we impose Robin boundary conditions.

\section{Bootstrapping hydrogen, revisited}
\begin{figure}[h!]
    \centering
    \includegraphics[scale = 0.35]{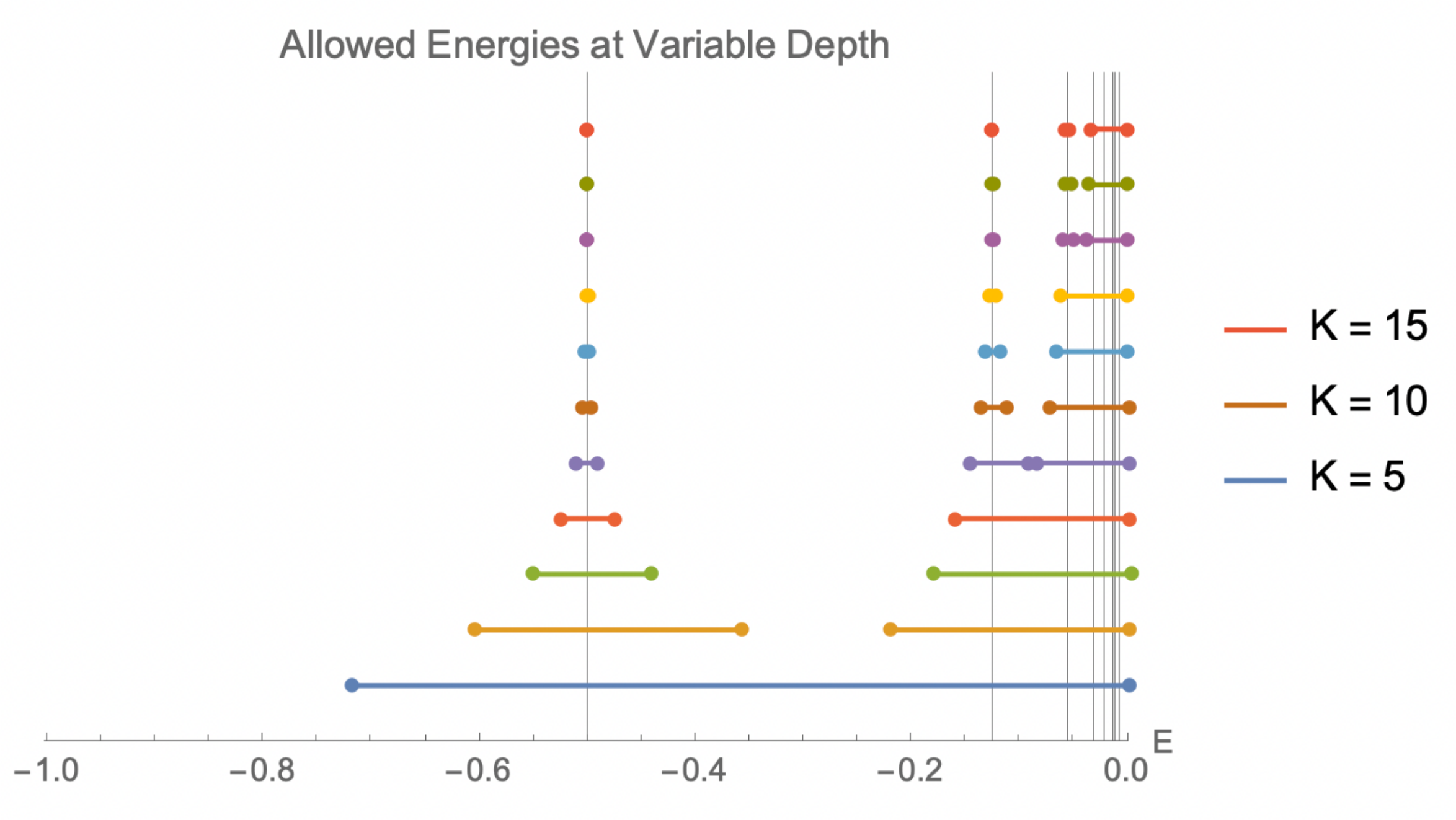}
    \caption{Allowed energies for the hydrogen bootstrap with $\ell = 0$, for sizes of Hankel matrices 5 - 15. Exact energies are in gray: in our units, they are $E_n = -1/(2n^2)$ for $n \geq \ell + 1$.}
    \label{fig:hbtsp}
\end{figure}

\begin{figure}[!h]
    \centering
    \includegraphics[scale = 0.4]{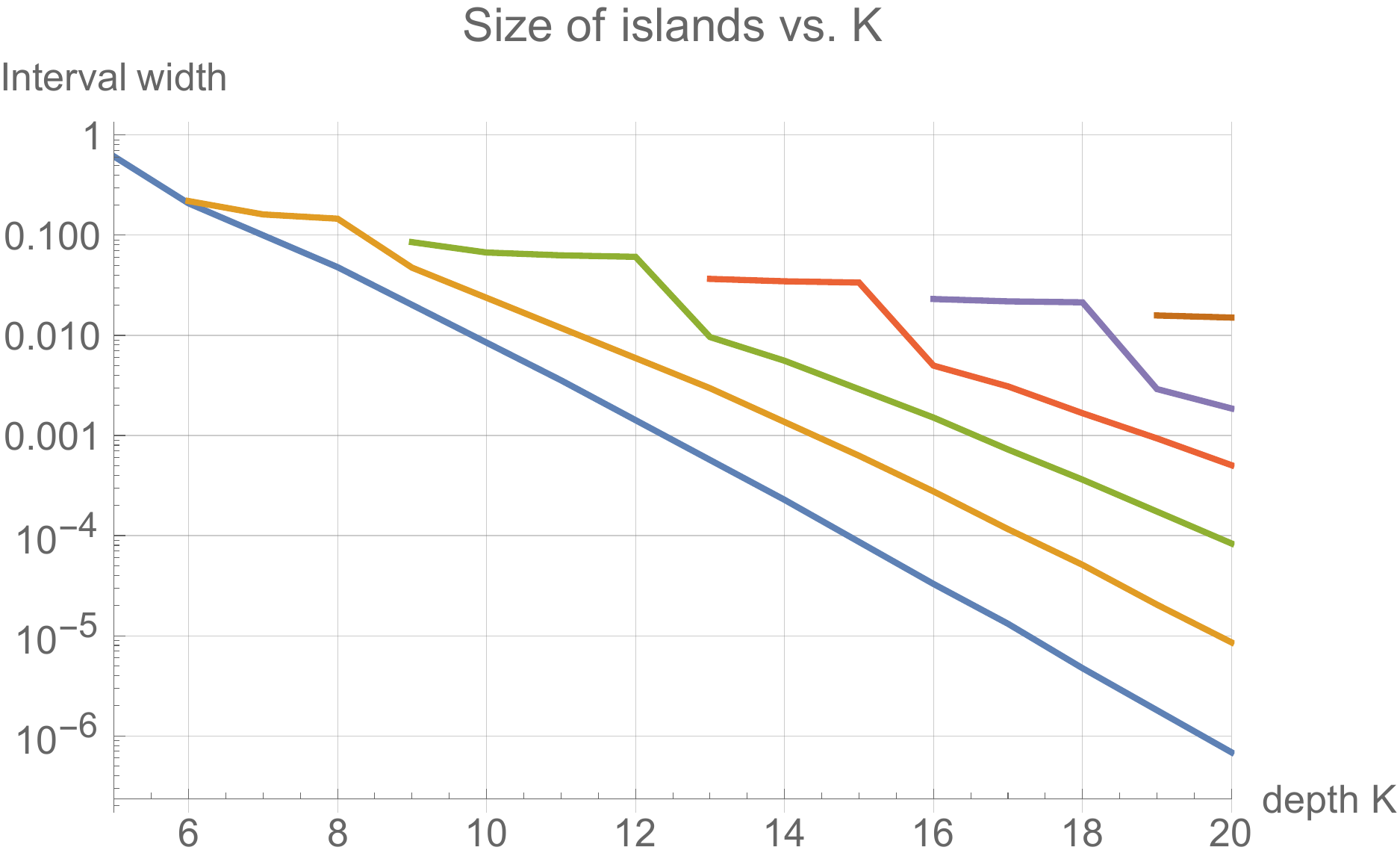}
    \caption{Interval width versus matrix size $K$ on a logarithmic axis, with $\ell = 0$. Each line represents an interval which forms around an exact energy level and shrinks as $K$ increases.}
    \label{fig:hconv}
\end{figure}

\begin{figure}[h!]
    \centering
    \includegraphics[scale = 0.4]{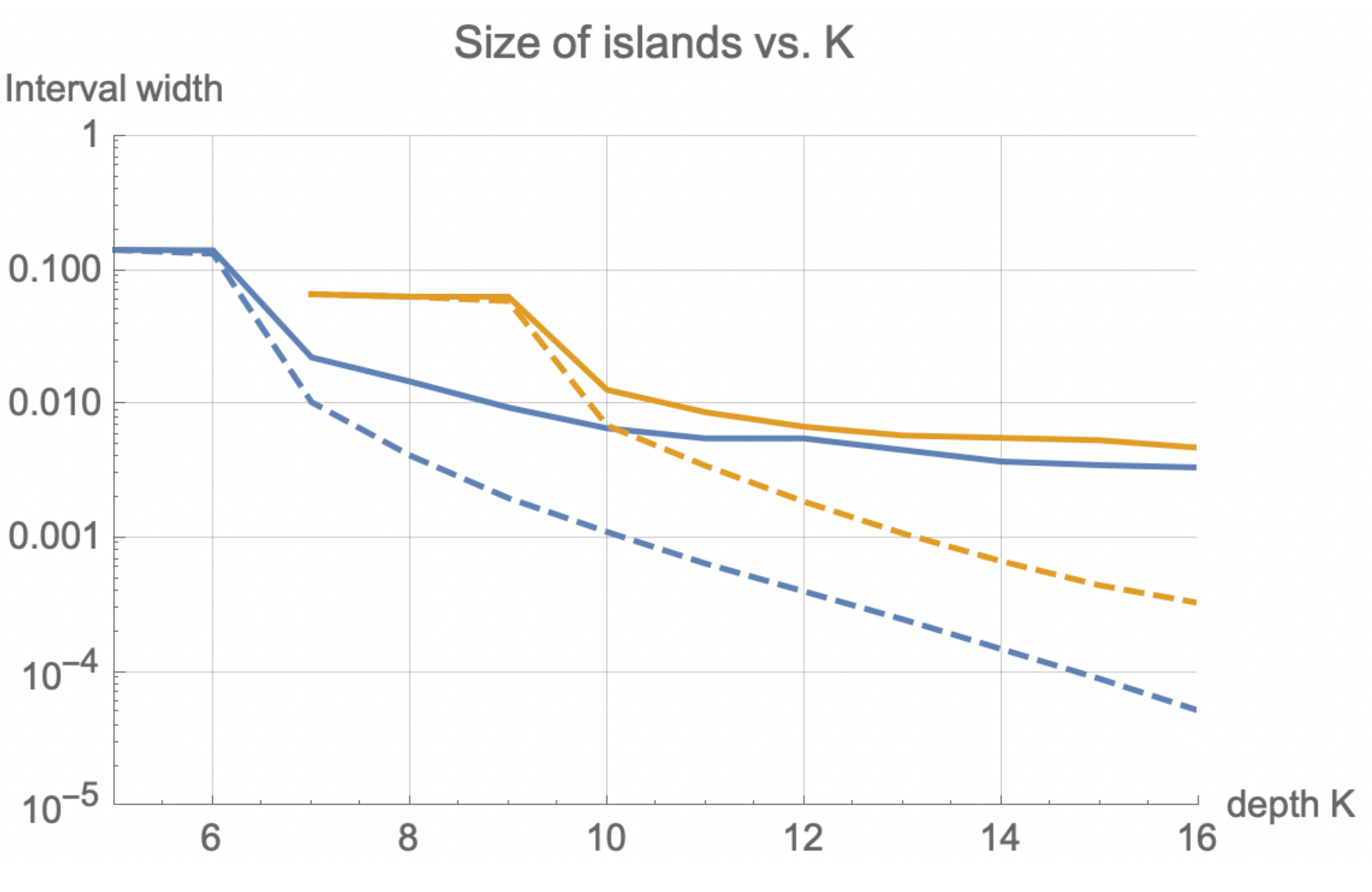}
    \caption{Comparing the Hamburger ($M \succeq 0$, solid) and Stieltjes ($M,\Tilde{M} \succeq 0$, dashed) checks. Interval width versus $K$ for the lowest two states, with $\ell = 1$. }
    \label{fig:hcomp}
\end{figure}

Here we present results from a numerical bootstrap of the hydrogen model, utilizing both Stieltjes matrices, instead of just the Hamburger matrix as was done in \cite{Berenstein:2021dyf}. We refer the reader to our previous paper for more background. To quickly summarize, candidate values of the energy $E$ of some eigenstate are chosen from an interval. The following recursion relation between moments $\langle r^n\rangle$ holds for energy eigenstates:
\begin{equation}\label{eq:hrec}
    0=8 m E\left\langle r^{m-1}\right\rangle+(m-1)[m(m-2)-4 \ell(\ell+1)]\left\langle r^{m-3}\right\rangle+4(2 m-1)\left\langle r^{m-2}\right\rangle
\end{equation}
Thankfully this recursion may be initialized only with the energy of the state $E$, which by the virial theorem directly determines the moment $\er{-1}$. We choose values of the energy $E$, generate a moment sequence of some length, and apply the positivity conditions of the Stieltjes moment problem for a matrix of finite size $K$. This allows us to rule out energy values which do not correspond to eigenstates. The result, for different sizes $K$ of the pair of Hankel matrices, is shown in Fig. \ref{fig:hbtsp}. 
The essential behavior of the bootstrap algorithm is the same. The convergence using both Stieltjes matrices is exponential with a speedup over using just the Hamburger matrix. We can also easily detect the $\ell = 0$ states, which were previously inaccessible. Fig. \ref{fig:hbtsp} shows qualitatively how the allowed intervals converge. For intervals which form around a given energy level, we can plot the convergence with $K$ on a logarithmic plot and see that it is exponential in the matrix size $K$, as in Fig. \ref{fig:hconv}. The flat portions which begin each curve represent the depths before which a given interval becomes disjoint. 
The addition of the second Stieltjes matrix is crucial for getting the bootstrap to pick up the $\ell = 0$ states. Including this matrix improves our earlier results considerably: the spectrum is completely detectable and the convergence is  better, especialy for low $\ell$ as show in Fig. \ref{fig:hcomp}.

\subsection{$\ell < 1$ and strange states}
An outstanding question about the quantum mechanical bootstrap is what data it truly receives about the problem. For instance, the bootstrap is completely agnostic about the quantization of the angular momentum parameter $\ell$, provided one forgets about the three-dimensional origin of the model. A spectrum exists for the radial Hamiltonian for any (positive) value of the parameter $\ell$. We want to analyze this possibility more carefully to try to understand how the bootstrap deals with this problem. There are two reasons to do this. First, the problem is analytically soluble. Secondly, 
the solutions can become slightly singular at $r=0$; this can be used to better understand what happens at the boundary of the interval and how the bootstrap program responds to that information. 

\subsubsection{Analytical predictions}
When the azimuthal parameter $\ell$ is quantized, the solutions to the radial hydrogen problem are Laguerre polynomials with exponential decorating factors. Let us relax the quantization condition and consider the equation
\begin{equation}\label{eq:hdiff}
    -\frac{1}{2}f''(r) + \left[\frac{\ell(\ell+1)}{2r^2} - \frac{1}{r}\right]f(r) = (-E)f(r)
\end{equation}
for arbitrary $0 < \ell < 1$ and with $r > 0, E > 0$. Multiplying this by $r^2$ brings it into a form similar to that of the hypergeometric differential equation. The general solution is expressed in terms of Whittaker's confluent hypergeometric functions: 
\begin{equation*}
    f(r) = \alpha M_{k,\mu}(z) + \beta M_{k,-\mu}(z)
\end{equation*}
where the parameters are:
\begin{equation*}
     k = \frac{1}{\sqrt{2E}} \qquad \mu^2 = (\ell + \frac{1}{2})^2 \qquad z = 2r/k  
\end{equation*}
Another set of solutions is given by the Whittaker $M$ and $W$ functions, but the basis above will work well for our purposes. 

We require that the solution $f(r)$ is in $L^2(\Rp)$. This requires that solutions vanish at infinity. For real $z \to \infty$, the Whittaker $M$-function has leading order asymptotic expansions \cite{dlmf}
\begin{equation}
    M_{k, \pm\mu}(z) \sim \frac{\Gamma(1\pm2 \mu)}{\Gamma(\frac{1}{2}\pm\mu-k)} \mathrm{e}^{\frac{1}{2} z} z^{-k} 
\end{equation}
The $M$-function diverges exponentially at infinity unless the gamma function in the denominator diverges as well. This would require
\begin{equation}\label{eq:qr1}
    \frac{1}{2} \pm \mu - k  = -n\quad \text{ where }\quad n \in \mathbb{N}
\end{equation}
Keeping this in mind, we can examine the behavior of these functions near the origin. Not all functions in the Hilbert space are finite at 0---they need only be normalizable. This means that we may have $f(r) = c z^s[ 1 + \OO(z)]$ for $s > -1/2$ and still have a function which is locally $L^2$ at the origin. As $z \to 0$, the $M$-function behaves as
\begin{equation}
    M_{k, \pm\mu}(z)=z^{\frac{1}{2}\pm\mu}[1+\OO(z)]
\end{equation}
For $\mu > 0$ the $(+)$ branch is zero at the origin and is acceptable. If $1/2 < \mu < 1$, the $(-)$ branch is square-integrable but infinite at $r = 0$, which is also acceptable. But if $\mu \geq 1$, the $(-)$ branch will be non-normalizable.

The condition \eqref{eq:qr1} is exactly a quantization condition. Let us consider first the $(+)$ branch. Rewriting in terms of physical parameters, it says that
\begin{equation*}
    1 + \ell + n = k = \frac{1}{\sqrt{2E}}
\end{equation*}
for some non-negative integer $n$. The physical energy is $E_{ph} = -E$ and is thus quantized by principal number $n > 0$ as
\begin{equation}\label{eq:qr}
    E^{(+)}_{ph} = -\frac{1}{2(n+\ell + 1)^2}
\end{equation}
which is exactly the same as the quantization rule for integral $\ell$, simply continued to fractional values (note that $n$ now starts at 0). For the $(-)$ branch of \eqref{eq:qr1}, we find the quantization rule
\begin{equation}\label{eq:qrminus}
    E^{(-)}_{ph} = -\frac{1}{2(n-\ell)^2}
\end{equation}
for $n \geq 0$. Recall that this only corresponds to normalizable eigenfunctions in the regime $0 < \ell < 1/2$. Despite their normalizability, they are infinite at the origin. As a result there are no inverse radial moments $\er{-p},\ p>0$ which are defined for these solutions. 

\subsubsection{Bootstrapping $0 < \ell < 1$}
\begin{figure}[h!]
    \centering
    \includegraphics[scale = 0.35]{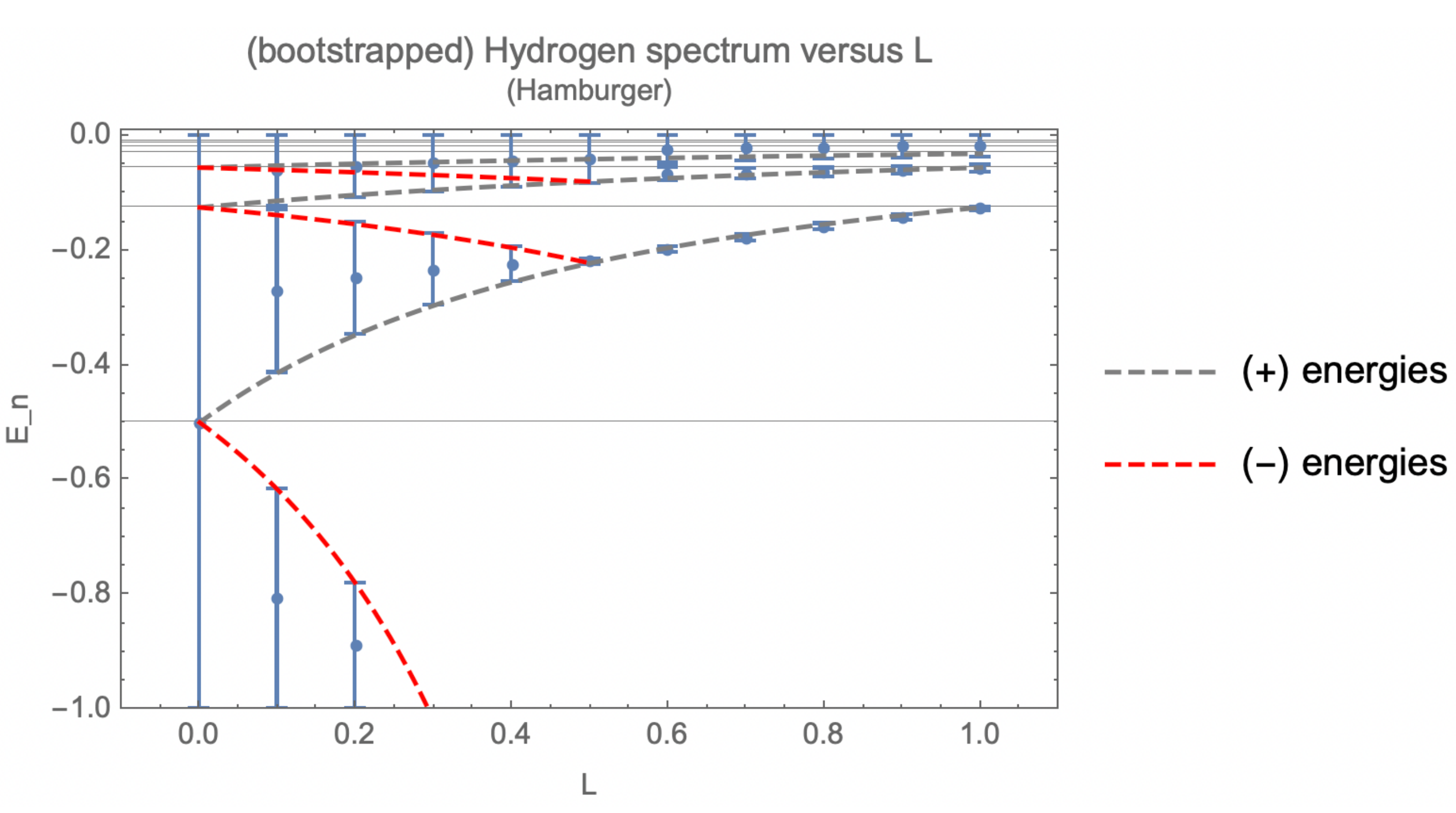}
    \caption{$K = 10$ bootstrap for various values of $\ell$, using only the matrix $M_{ij}= \langle r^{i+j}\rangle$. One can see that convergence grows better as $\ell$ increases. Exact (hydrogen) energies in gray; Whittaker predictions in dashed gray, red. }
    \label{fig:hspecm1}
\end{figure}

Running a bootstrap for values $0 \leq \ell \leq 1$ gives an interesting regime in which to examine how the two Stieltjes matrices affect convergence and to see the signatures of the Whittaker functions. Fig. \ref{fig:hspecm1} displays bootstrap data for $K = 10$ at various values of $
\ell$, showing the allowed energy intervals vertically and checking only the Hamburger matrix $M_{ij} = \langle r^{i+j}\rangle$. 
When $\ell = 0$, checking the Hamburger matrix alone does not disallow any energy values. As $\ell$ increases to fractional values, the allowed intervals shrink and the positions of the ``excited" intervals shift upwards, in accordance with perturbative expectations. Once $\ell = 1$ the Hamburger matrix works decently to
 bootstrap all the excited states. It should be noted that at $\ell=1/2$ it is well known that the Hamiltonian becomes essentially self-adjoint: this coincides with the disappearance of the second branch of solutions (the ones that become non-normalizable). This evidence makes it plausible that to understand the issues that arise due to the boundary, we need to look very carefully at the question of which operators are self-adjoint. 

When we run the same experiment but use the positivity checks for both Stieltjes matrices (i.e. now including $\Tilde{M}_{ij} = \er{1+i+j}$), some interesting results emerge. First, the $\ell = 0$ states appear. There is also a new, disjoint series of intervals that the bootstrap detects which decrease in energy as $\ell$ increases. 
\begin{figure}[h!]
    \centering
    \includegraphics[scale = 0.35]{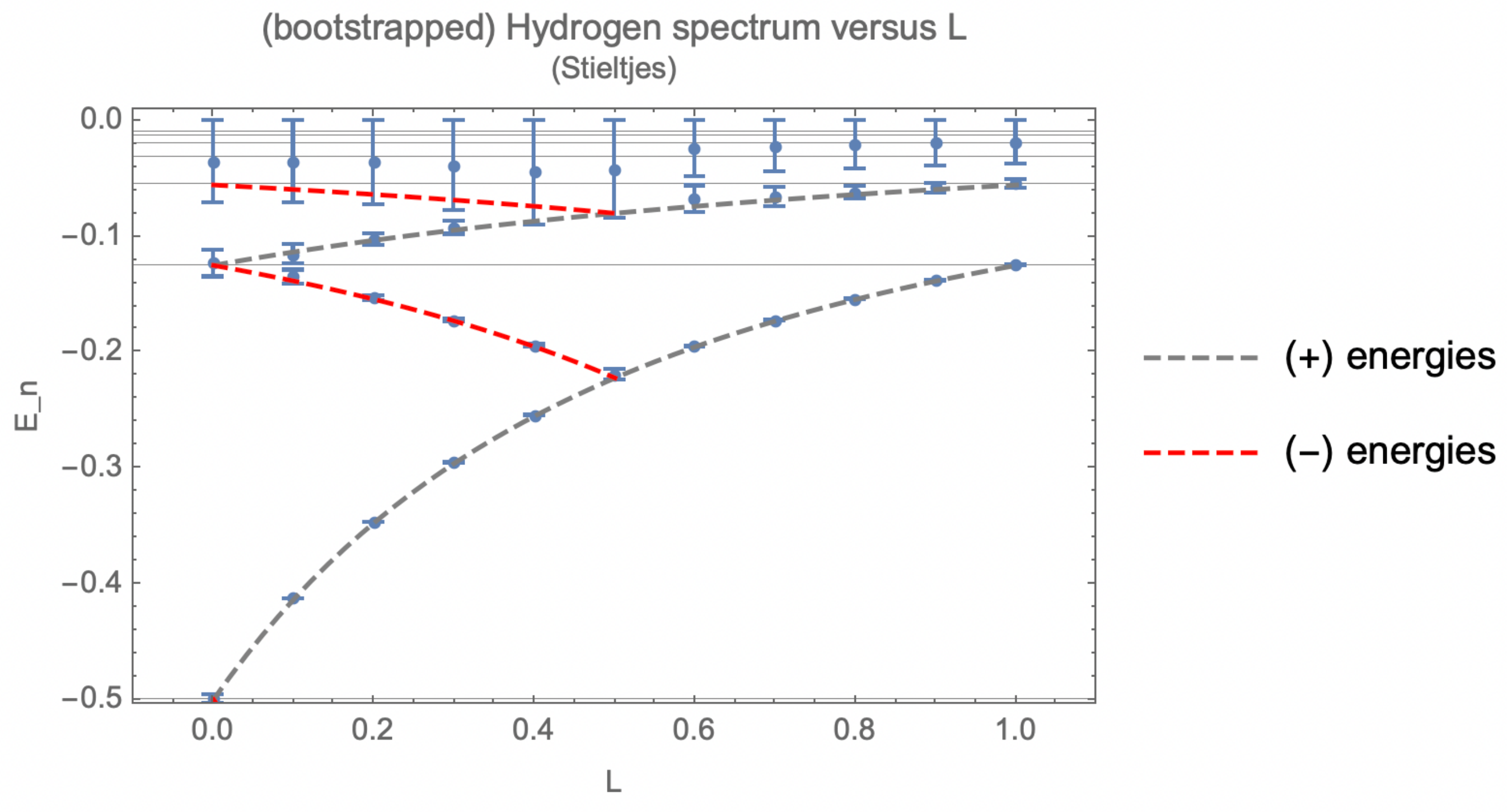}
    \caption{$K = 10$ bootstrap with both Stieltjes matrices $M,\Tilde{M}$. There is a set of ``states" which decrease in energy as $\ell$ increases, only while $0 \leq \ell \leq 1/2$.}
    \label{fig:hspecm2}
\end{figure}
These are precisely the states with energies \eqref{eq:qrminus}, which are infinite at the origin. We can eliminate them by adding another matrix to our positivity constraints. 

Consider the matrix with elements $M'_{ij} = \langle r^{i+j-1}\rangle, 0 \leq i,j \leq K-1$. For any eigenstate accessible through the recursion, the $\langle r^{-1}\rangle$ moment is well-defined and proportional to the energy of the state by the virial theorem (this is implied by e.g. \eqref{eq:hrec} with $m = 1$). Positivity of $M'$ is thus another necessary condition for moment sequences derived from physical (finite energy) eigenstates of the hydrogen Hamiltonian \footnote{The physical requirement here is that e.g. the moment $\langle V \rangle_\psi$ is well-defined for eigenstates $\psi$; this is not mathematically required for the pure eigenvalue problem.}.

\begin{figure}[h!]
    \centering
    \includegraphics[scale = 0.35]{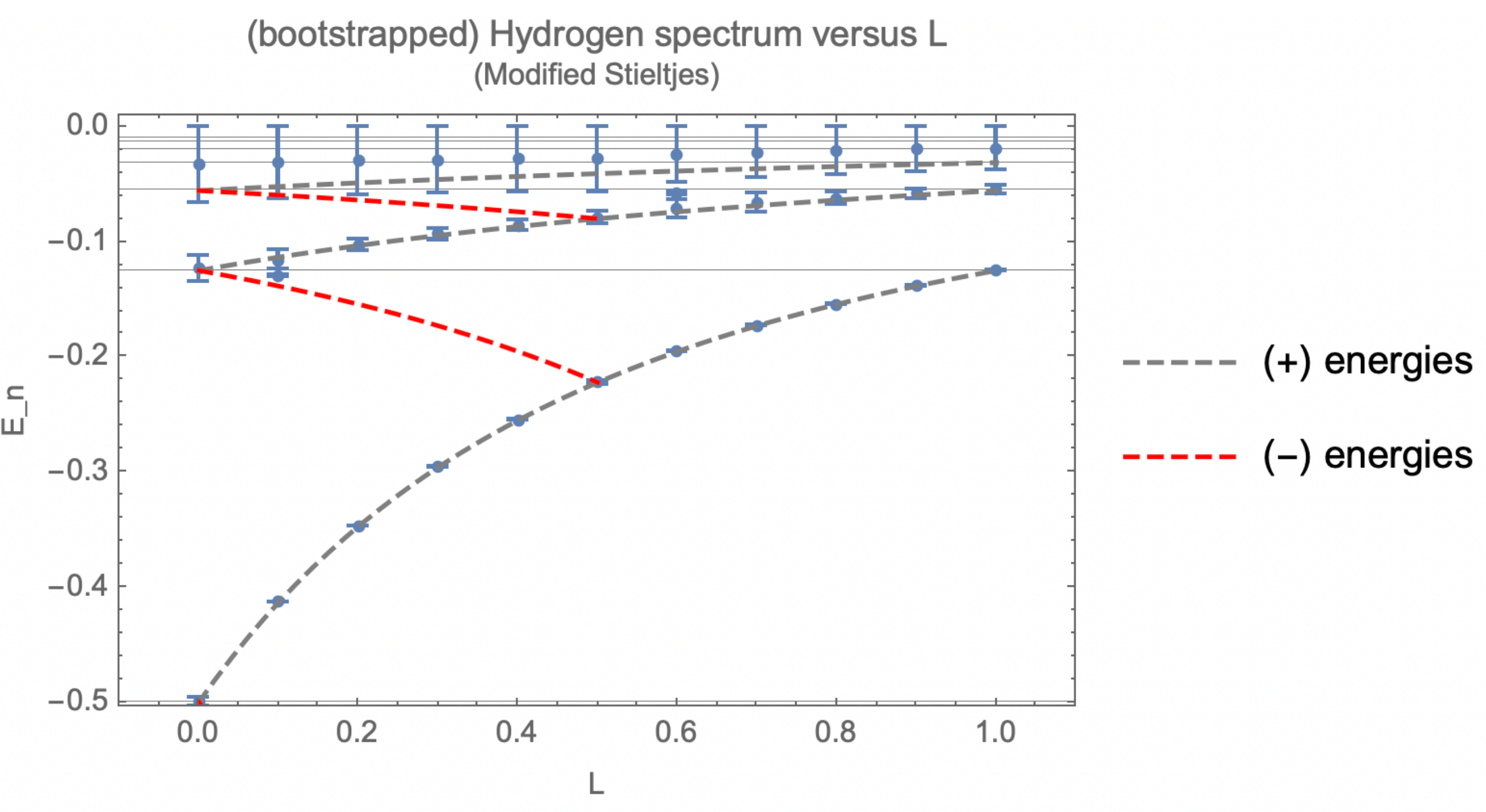}
    \caption{$K = 10$ bootstrap checking the three matrices $M, \Tilde{M},M'$. Spectrum flows upward as $\ell$ increases, as expected from perturbation theory. The bootstrap now does not detect the states \eqref{eq:qrminus}.}
    \label{fig:hspecm3}
\end{figure}

Finally, we can carry out a bootstrap where we check positivity of both Stieltjes matrices $M, \Tilde{M}$ in addition to the matrix $M'$ just introduced. Shown in Fig. \ref{fig:hspecm3}, adding this additional positivity constraint eliminates the descending states visible for $\ell < 1/2$. 
While demanding positivity of this final matrix $M'$ is reasonable within the context of quantum mechanics, from the Stieltjes problem point of view these states did correspond to acceptable probability measures. By enforcing that the first inverse moment $\er{-1}$ is defined, we were able to impose a ``soft" boundary condition on the state. 
\section{Fixing the microcanonical bootstrap}
In correctly implementing the bootstrap for the hydrogen model, we learned that including new positivity checks carved out new regions of allowed parameter space at a given depth $K$. For example, with $\ell = 0$, using only the Hamburger matrix left a large region of parameter space (energy) unconstrained. 

\begin{figure}[h!]
    \centering
    \includegraphics[scale = 0.6]{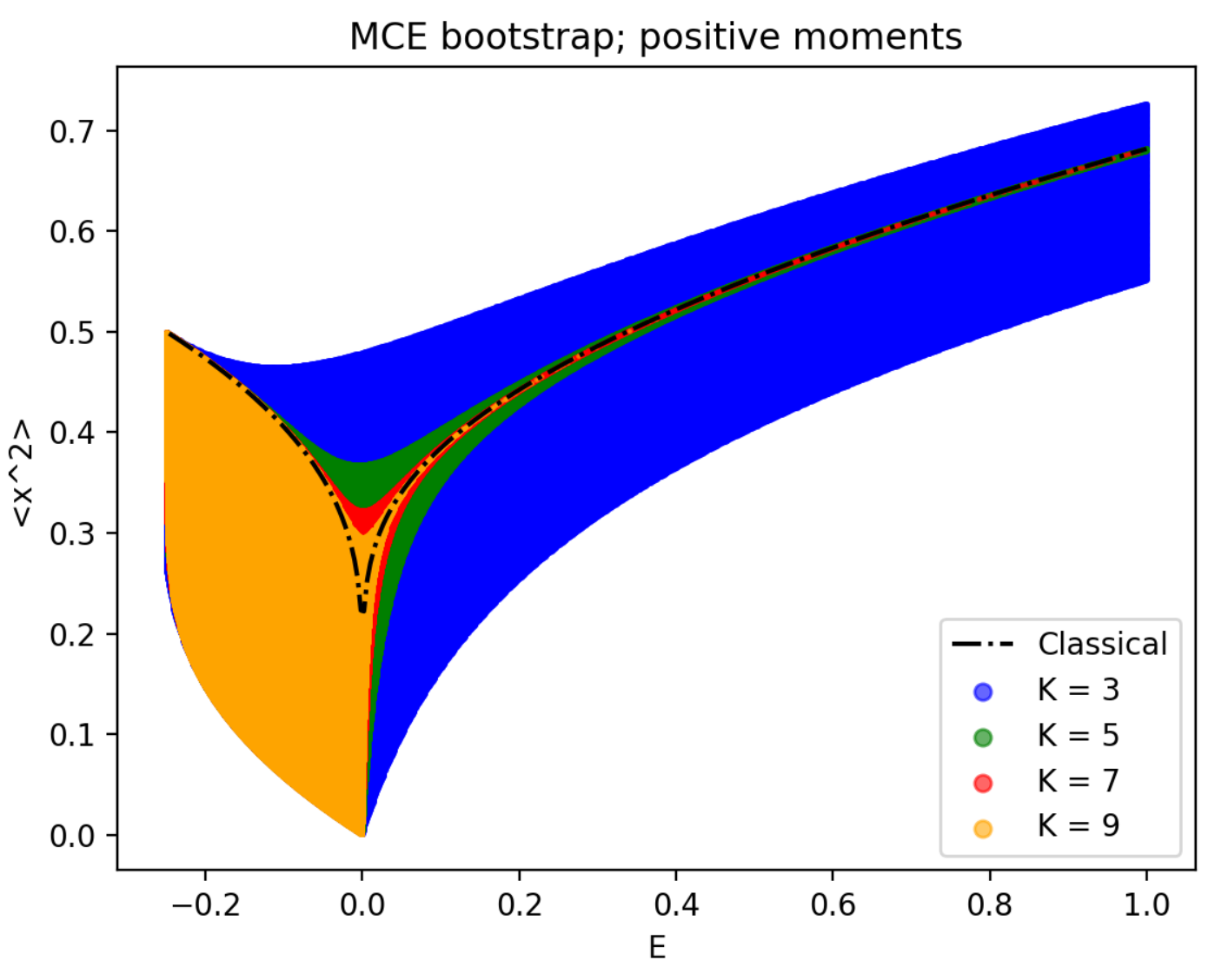}
    \caption{Allowed parameter values for various depths $K$ of the Hamburger matrix and classical relation pictured. For $E < 0$ a large `peninsula' appears.}
    \label{fig:mce_pos}
\end{figure}

Recently, Nakayama \cite{Nakayama:2022ahr} has explored the idea of bootstrapping the microcanoncial ensemble (MCE) of a given classical dynamical system. This is exactly the $\hbar \to 0$ limit of a quantum system: normalizabilty and probabilistic interpretations remain but the dynamics are altered. Specifically, one term of the recursion (a term that is  proportional to $\hbar$) vanishes. The recursion for moments of measures on $\RR$ was
\begin{equation}\label{eq:oldrec}
     0=2 m E \langle x^{m-1} \rangle+ \frac{1}{2}m(m-1)(m-2)\ex{n-3}-\langle x^{m} V^{\prime}(x) \rangle-2 m \langle x^{m-1} V(x) \rangle
\end{equation}
The MCE moment recursion for a general potential on $\RR$ is
\begin{equation}\label{eq:mce_rec}
    0=2 m E \langle x^{m-1} \rangle- \langle x^{m} V^{\prime}(x) \rangle-2 m \langle x^{m-1} V(x) \rangle
\end{equation}
Nakayama considers this for the double-well potential $V(x) = -x^2 + x^4$. They perform a numerical bootstrap checking positivity of the Hamburger matrix. Demanding that $\ex{} = 0$ constrains all odd moments to vanish. The Hamburger matrix thus takes the form
\begin{equation*}
    M^{(K)}=\left[\begin{array}{ccccc}
1 & 0 & \ex{2} & \cdots  & \ex{K-1} \\
0 & \ex{2} & &  & \vdots \\
\ex{2} & & \ddots & & \vdots \\
\vdots & & & \ddots & 0 \\
\ex{K-1} & \cdots & \cdots & 0 & \ex{2 K-2}
\end{array}\right]
\end{equation*}
The result of checking positivity of this matrix in the $\{E,\ex{2}\}$ plane is shown in Fig. \ref{fig:mce_pos}.
The result is that for energies $E > 0$, which live above the double well, checking positivity of the Hamburger matrix works well. At successive depths the allowed region of parameter space shrinks to a small envelope around the classical $E, \ex{2}$ curve. This was the result of Nakayama, who noted that the large `peninsula' for $E < 0$ persists at higher depths of the Hamburger matrix. They conjectured that the peninsula was a feature of the MCE bootstrap, not a bug.

This would be suprising: the expectation should be that the exact allowed region ($K \to \infty$) in the MCE bootstrap is exactly the classical curve relating $E, \ex{2}$---since every member of the ensemble is just the classical system. This is apparently not the result obtained using just the Hamburger matrix.

However, there is a large family of constraints that we are missing! For any energy $E < 0$, the particle spends no time at the origin. In other words, given an energy $E < 0$, the associated classical motion has zero support as $x \to 0$ in phase space. As a result, all (classical) inverse moments of $x$ are finite. That is to say that when $E <0$, the following integral over the classical motion converges for $\forall n \in \mathbb{Z}$:
\begin{equation*}
    \ex{n}_{cl} = \frac{1}{T}\oint \frac{x^n}{\sqrt{E +x^2 - x^4}}\ dx
\end{equation*}
where the integral here is over an orbit of the classical motion defined by the turning points $x_i:-x_i^2 + x_i^4 = E$. Since all the inverse moments are well-defined, we may consider operators $\OO_I = \sum_{n=-K}^Kc_nx^n$ acting on states. This will descend to a positivity constraint on a new Hankel matrix $M_I$ which contains inverse as well as positive moments. The matrix $M_I$ at level $K$  will take the form
\begin{equation*}
    M^{(K)}_I=\left[\begin{array}{ccccc}
\ex{-(K-1)} & \ex{-(K-2)} & \cdots & \ex{-1}  & 1 \\
\ex{-(K-2)} & \ex{-(K-3)} & &  & \ex{} \\
\vdots & & \ddots & & \vdots \\
\ex{-1} &  & & \ddots & \ex{K-2} \\
1 & \ex{} & \cdots & \ex{K-2} & \ex{K-1}
\end{array}\right]
\end{equation*}
Norm positivity $\langle \OO_I^\dagger \OO_I \rangle \geq 0$ implies that $M^{(K)}_I \succeq 0$ for all depths $K$. This becomes an additional positivity constraint which is well-defined for $E<0$. 

\begin{figure}[h!]
    \centering
    \includegraphics[scale = 0.6]{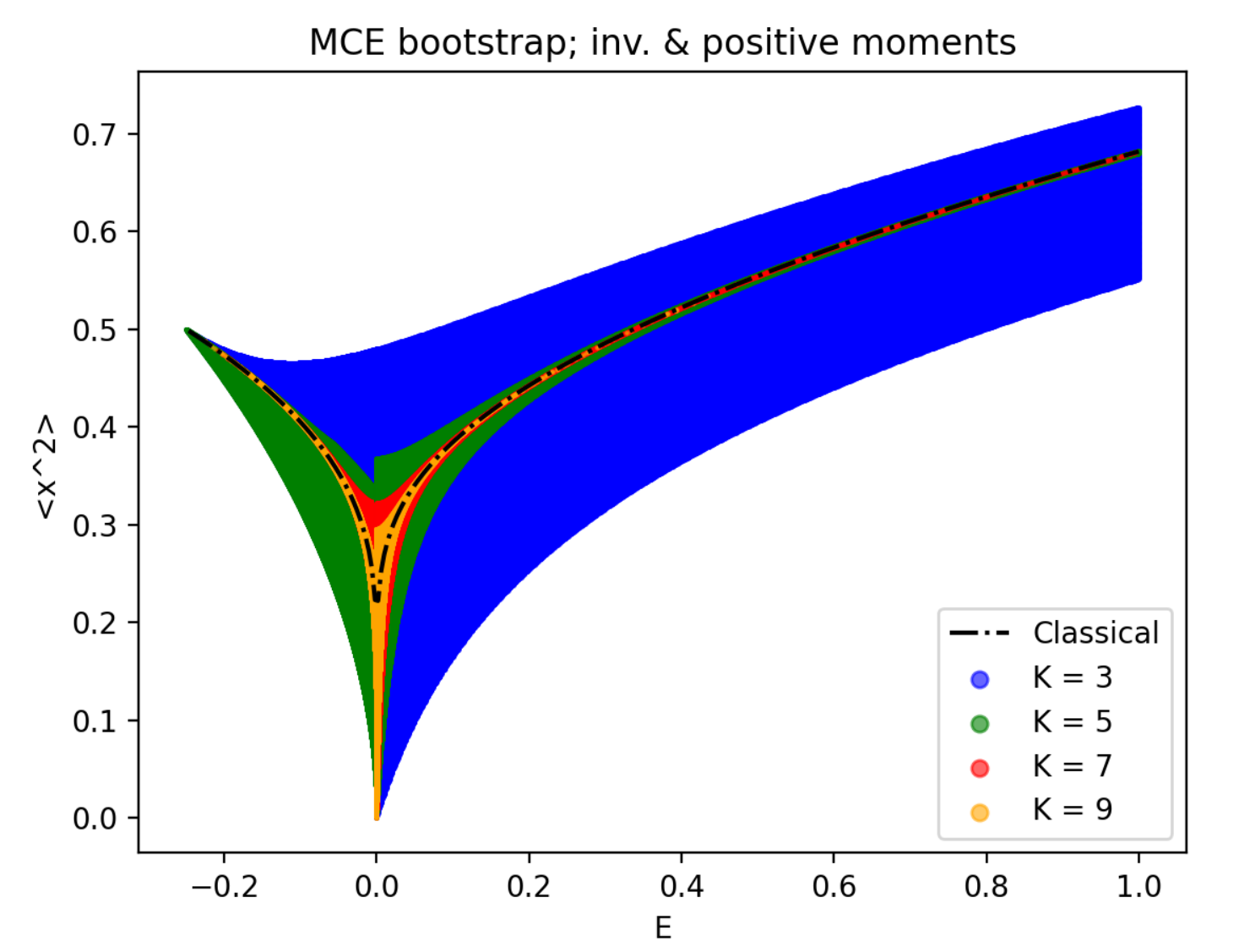}
    \caption{`Fixed' MCE bootstrap for various depths $K$. Including the new positivity constraint removes the `peninsula' and the bootstrap converges everywhere to the classical curve.}
    \label{fig:mce_both}
\end{figure}

To integrate this into the bootstrap, we simply use the recursion \eqref{eq:mce_rec} to generate negative moments, using the parameters $E, \ex{2}$ to initialize as before. Then we can carry out a bootstrap checking only the Hamburger positivity constraint for energies $E>0$ as before, but checking positivity of both the Hamburger matrix and the new matrix $M_I$ for energies $E < 0$. The result, shown in Fig. \ref{fig:mce_both}, conforms to expectations: the bootstrap converges everywhere to a small envelope surrounding the classical curve.
This simply shows that to get the bootstrap to work properly, one needs to consider all  physically allowable constraints. By including these inverse moments one can cut down on the peninsula of allowed parameter space. 
In this case, when we test for the inverse moments we are checking that the particle does not reach into the forbidden region. The Hamburger problem would tell us that for each of the unphysical solutions there is still a measure that satisfies 
the moment problem. That measure must necessarily be non-vanishing near the origin: otherwise the inverse moments would be well defined. 
Such a measure that does not vanish near the origin would violate conservation of energy in the classical system, where there can not be any tunneling.
The lesson here is clear, as was also the case for the hydrogen atom: the classical (mathematical) moment problem alone is not sufficient to determine completely the physically acceptable solutions. 

Additional constraints might be required that enlarge the set of inequalities to test. Only when this additional input is specified do we get a complete solution. This should be contrasted with the statement found in  footnote (20) of \cite{Han:2020bkb}, which implicitly argues that the procedure always converges (defines a density matrix), but where the list of {\em all} operators ${\cal O}$ described in that paper is not sufficiently detailed to guarantee convergence to the correct answer.

\section{Bootstrapping the Half Line}
\begin{figure}[h!]
    \centering
    \includegraphics[scale = 0.4]{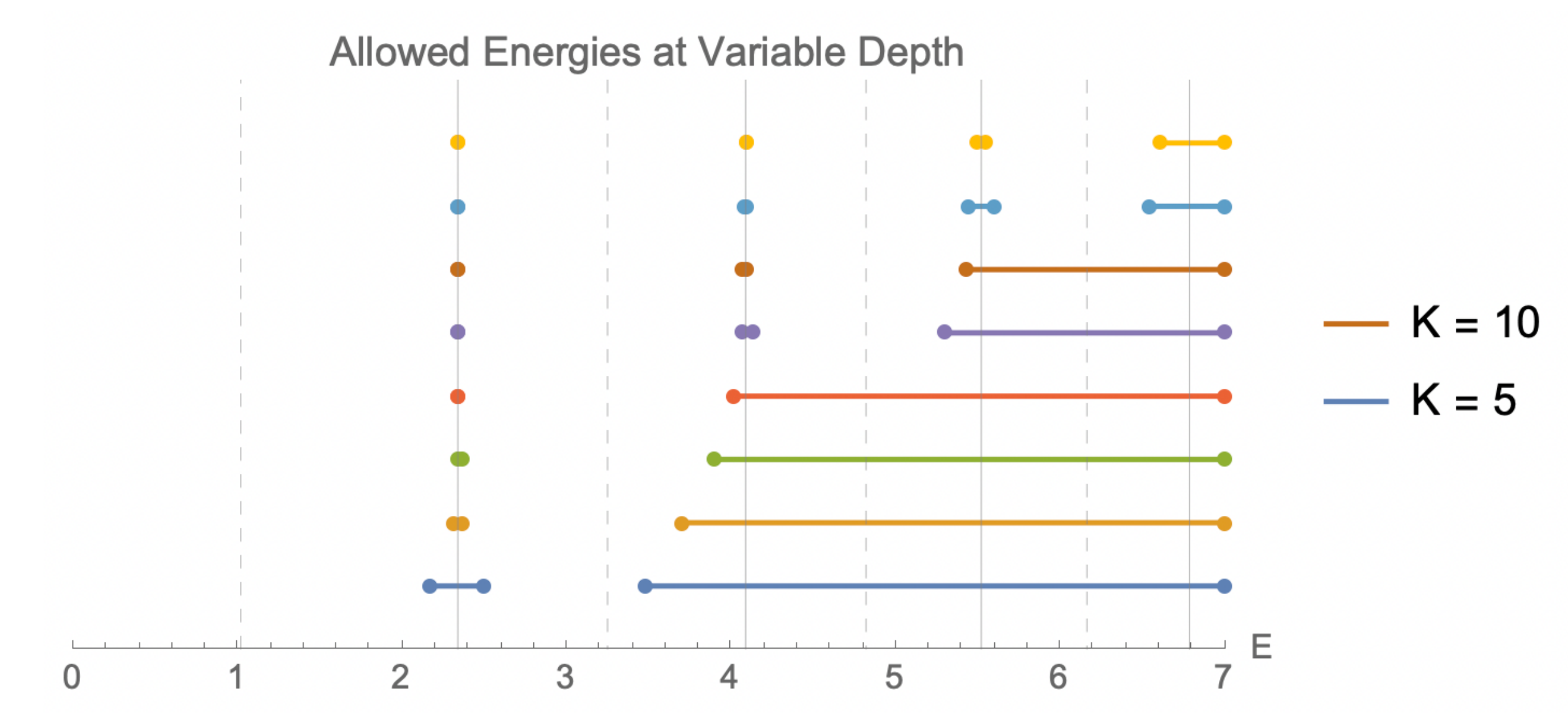}
    \caption{Numerical bootstrap for depths $5 \leq K \leq 12$ (increasing vertically) for the Airy model/linear potential. Intervals are energy values allowed at a given depth. Dashed vertical lines are the `exact' energies for Neumann ($\psi'(0) = 0$) states and solid vertical lines are the Dirichlet ($\psi(0) = 0$) energies.}
    \label{fig:airyspec}
\end{figure}

The hydrogen problem showed that bootstrapping on the half line is not quite the same as bootstrapping on $\RR$. Let us consider another problem from undergraduate quantum mechanics: the linear potential. This model is nice because like hydrogen, the recursion for the model may be initialized by the energy $E$ alone:
\begin{equation}\label{eq:airyrec}
    \left\langle x^{m}\right\rangle=\frac{1}{2 m+1}\left[2 m E\left\langle x^{m-1}\right\rangle+\frac{1}{2} m(m-1)(m-2)\left\langle x^{m-3}\right\rangle\right]
\end{equation}
We consider this problem on the half line, and carry out an algorithm identical to that of the hydrogen bootstrap, using the Stieltjes positivity check. The result is a bootstrap which converges nicely to the exact energies as computed by standard numerical techniques, as in Fig. \ref{fig:airyspec}. Notably, the bootstrapped spectrum corresponds to the exact energies of the system with \textit{Dirichlet} boundary conditions for the wavefunctions: $\psi(0) = 0$. Why has the bootstrap selected this boundary condition versus a mixed or Neumann condition? After all, when dealing with the trigonometric bootstrap on the circle \cite{Aikawa:2021eai,Tchoumakov:2021mnh,Berenstein:2021loy}, all the possible quasiperiodic boundary conditions appeared as possible solutions of the bootstrap equations. Essentially, the recursion relations were agnostic in that case on the specific choice of boundary conditions.

The answer in this case is that because we have failed to specify boundary conditions at the origin, the recursion \eqref{eq:airyrec} is incomplete (data is missing).
 The issue of boundary conditions in the bootstrap is subtle.  The half line provides a good testing ground for dealing with the boundary conditions, as it is 
really not \textit{a priori} clear what to do about the origin. 

It turns out that the half line is rife with issues as a quantum mechanical system. We will introduce these issues and see that they are essentially related to questions of the domains of certain unbounded operators. In order to find a self-adjoint Hamiltonian, we will need to define some boundary conditions. We will see how these boundary conditions enter the bootstrap recursion relations, then revisit the linear potential. 

\subsection{Quantum mechanics on the half line}
For simplicity, we will assume that ``doing quantum mechanics" on the half line means solving the eigenvalue problem
\begin{equation}\label{eq:tise}
    -\frac{d^2\psi}{dx^2} + V(x)\psi = E\psi 
\end{equation}
for functions $\psi(x) \in L^2(\RR_+)$ on the half line $x \geq 0$ and their eigenvalues $E$. In the usual quantum mechanical treatment, one would say that the Hamiltonian is the operator
\begin{equation}\label{eq:halflineH}
    \hat{H} = p^2 + V(x)
\end{equation}
where the operators $x,p$ obey the canonical relation $[x,p] = i$. We wish to determine the spectrum of $H$. Of course, to serve as the physical energy operator of a system, the Hamiltonian should have only real eigenvalues; it must be self-adjoint. Textbooks would introduce this as the condition that $(\phi,H\psi) = (H\phi,\psi)$ for all states $\psi,\phi \in \HH$. In a finite-dimensional Hilbert space, this is the condition of `Hermiticity', and is equivalent to self-adjointness. But in the infinite-dimensional case we are interested in, the situation is more subtle. 

Most operators in familiar one dimensional quantum mechanics are unbounded---indeed, any Hamiltonian with arbitrarily large eigenvalues is unbounded. This means that for $\psi \in \HH$, an unbounded operator $A$ may map states out of the Hilbert space, i.e. $A\psi \notin \HH$. To avoid this possibility, the definition of an unbounded operator $A$ consists of how the operator acts on functions as well as a declaration of an operator domain $D(A)$, a dense subspace of $\HH$. The domain $D(A)$ is the preimage of $\HH$ under the operator $A$.  

The required restriction of operator domains is just the familiar task of supplying boundary conditions to eigenvalue problems like \eqref{eq:tise}. By supplying boundary conditions for the solutions, we eliminate some functions in the Hilbert space from the domain of consideration for the operator. The role of these operator domains has been extensively studied; see e.g. \cite{reed1,reed2,anomaly,anomaly1}. 

Operator domains are also important for self-adjointness. Two conditions must be satisfied for an unbounded operator $H$ with domain $D(H)$ to be self-adjoint. It must be \textit{symmetric}, i.e.
\begin{equation*}
    (H\phi,\psi) = (\phi,H\psi) \qquad \forall \phi,\psi \in D(H) \subset \HH
\end{equation*}
The second condition is that its adjoint shares the same domain: $D(H) = D(H^\dagger)$. In general, $D(H) \subseteq D(H^\dagger)$. Only when the domain of the operator coincides with the domain of its adjoint does the spectral theorem apply. We will consider two examples of these issues: the momentum operator on the half line and the Hamiltonian \eqref{eq:halflineH} on the half line. 

\subsubsection{No momentum on the half line?}
Let us consider the operator $p = -i\partial_x$ acting on functions $\psi \in L^2(\Rp)$. The boundary conditions at infinity are already fixed by the Hilbert space. The operator $p$ is symmetric when $(p\phi,\psi) - (\phi,p\psi) = 0$. This places conditions on the domain $D(p)$:
\begin{equation*}
    (p\phi,\psi) - (\phi,p\psi) =0 =  \intplus i\Bar{\phi}'\psi +i \Bar{\phi}\psi' = i\Bar{\phi}(0)\psi(0)
\end{equation*}
where we have integrated by parts. This seems to suggest that we define $D(p) = \{\psi \in L^2(\Rp) \ |\ \psi(0) = 0\}$, so that $p$ is symmetric on $D(p)$. But what about $D(p^\dagger)?$ By definition the adjoint satisfies $(p^\dagger \phi,\psi) = (\phi,p\psi)$ for $\psi \in D(p)$ and $\phi \in D(p^\dagger)$. As above we can write
\begin{equation*}
    (p^\dagger\phi,\psi) - (\phi,p\psi) =  \intplus i\Bar{\phi}'\psi +i \Bar{\phi}\psi' = i\Bar{\phi}(0)\psi(0) = 0
\end{equation*}
Because of the conditions on $\psi$, the above is true for any $\phi \in L^2(\Rp)$ which is finite at the origin. Thus, $D(p^\dagger) \neq D(p)$, and the momentum operator is not self-adjoint on $D(p)$. 

One may wonder if we could suitably enlarge the domain of $p$ so that $D(p^\dagger) = D(p)$. In fact, there is a theory of such `operator extensions'. Given a symmetric, but not self-adjoint operator, it may be possible to extend the domain of definition so that the operator becomes self-adjoint. The existence of these self-adjoint extensions can be cleanly characterized in terms of deficiency indices: given a closed symmetric operator $A$, define two integers $n_\pm$ by
\begin{equation*}
    n_\pm = \dim\ \text{ker}[i\mp A^*]
\end{equation*}
Morally, this quantity measures `how much' of the spectrum of $A$ fails to be real, and hence how $A$ fails to be self-adjoint. It can be shown that  the operator $A$ is self-adjoint if and only if $n_\pm = 0$, and that $A$ has self-adjoint extensions if and only if $n_+ = n_-$ \cite{reed2}. By explicitly computing the deficiency subspaces, one can follow a maze of theorems to explicitly construct self-adjoint extensions. A useful theorem of von Neumann implies that for real potentials, the differential operator in \eqref{eq:tise} has equal deficiency indices $n_+ = n_-$.  

In the case of the momentum operator, one can easily solve the equations $p\psi = -i\psi' = \pm i\psi$ in $L^2(\Rp)$ and realize that the deficiency subspaces are mismatched: $n_+ \neq n_-$. This shows that there is no suitable self-adjoint momentum operator on the half line. Trying to define such an operator can lead to various paradoxes \cite{anomaly}. Recently, the authors in \cite{halflinemom} also considered these issues. They define a suitable momentum operator by passing to a cover of the Hilbert space. For our purposes, we are mostly concerned with the Hamiltonian, rather than the momentum alone.

\subsubsection{Self-adjoint Hamiltonians}
Despite not having a self-adjoint definition of momentum, we can define a domain on which the Hamiltonian \eqref{eq:halflineH} is truly self-adjoint. To construct this space we proceed essentially as before. The condition that the Hamiltonian $H = -\partial_x^2 + V(x)$ is symmetric is
\begin{equation*}
    (H\phi,\psi) - (\phi,H\psi) =0 =  \intplus -\Bar{\phi}''\psi + \Bar{\phi}\psi'' = \Bar{\phi}(0)\psi'(0) - \Bar{\phi}'(0)\psi(0)
\end{equation*}
This is satisfied if we require $\psi(0) + a\psi'(0) = 0$ (and similar for $\phi$) for a constant $a \in \mathbb{C} \cup \{\infty\}$ (these linear, mixed boundary conditions are sometimes called `Robin' conditions). Dirichlet conditions correspond to $a = 0$ and Neumann conditions to $a = \infty$. Let us consider this subset of $L^2(\Rp)$ as a candidate domain for $H$. Is $H$ self-adjoint on this domain? 

As above, we can calculate $(H^\dagger\phi,\psi) - (\phi, H\psi)$ for $\psi \in D(H),\ \phi \in D(H^\dagger)$ and demand that the result vanishes. Doing so gives the condition
\begin{equation*}
    [\Bar{\phi}(0) + a\Bar{\phi}'(0)]\psi'(0) = 0 \quad \implies \quad \phi(0) + \Bar{a}\phi'(0) = 0
\end{equation*}
This is exactly equivalent to the condition on the states $\psi \in D(H)$ provided the parameter $a$ is real. We can conclude that $H = -\partial_x^2 + V$ is self-adjoint on the domain 
\begin{equation}\label{eq:hamdoms}
    D_a(H) = \left\{ \psi \in L^2(\Rp)\ |\  \psi(0) + a\psi'(0) = 0 \quad a \in \RR \cup \{\infty\}\right\}
\end{equation}
We notice that there is a one-parameter family of such domains, indexed by the (dimensionful) extension parameter $a$. It remains to understand how confining ourselves to this domain will affect the recursion relations that generate the bootstrap. Indeed, the extension parameter $a$ does represent a physical aspect of the system to which the bootstrap should be sensitive!

To illustrate the physical consequences \cite{anomaly}, consider the free particle on $\Rp$; $-\psi'' = E\psi$ with $\psi(0) + a\psi'(0) = 0$. The solutions with $E = k^2$ are forward and backward traveling plane waves: 
\begin{equation*}
    \psi = Ae^{ikx} + Be^{-ikx}
\end{equation*}
When we impose the boundary condition, the solution becomes
\begin{equation*}
    \psi = A\left(e^{-ikx}+Re^{ikx}\right);\qquad R = \frac{aik-1}{aik+1}
\end{equation*}
We can interpret the (pure phase) $R$ as a reflection coefficient, $|R| = 1$. Physically, the interpretation is that the boundary conditions at the origin reflect waves with a phase shift $\arg R$ that depends on the extension parameter $a$.

Finally, we note that in the case of the moment problem/bootstrap on $\RR$, these issues of operator domain are of less concern. The boundary conditions associated with $L^2(\RR)$ ensure that most familiar Hamiltonians are essentially self-adjoint. This is detailed in appendix \ref{sec:apb}.

\subsection{Anomalies in the recursion}
We have seen that for problems on the half line, finding a self-adjoint Hamiltonian required us to define a one-parameter family of domains $D_a(H)$. The bootstrap recursion should be sensitive to this entire family of physically inequivalent quantizations of the system. 

To investigate these effects, let us consider the ``0+1" dimensional version of Noether's theorem, which is usually introduced as Ehrenfest's theorem \cite{sakurai}. For an operator $A$ and a Hamiltonian $H$, Ehrenfest's theorem governs the time evolution of expectation values of the operator $A$ in a state $\psi$:
\begin{equation}\label{eq:ehrenfest}
    \frac{d}{dt}\pex{A}  = \left\langle \frac{\partial A}{\partial t}\right\rangle_\psi + i \pex{[H,A]}
\end{equation}
When we derived the bootstrap recursion (e.g. in \cite{Han:2020bkb, Berenstein:2021dyf}), we used the linear constraints $\pex{[H,\OO_i]} = 0$ for some set of operators $\OO_i$, which are usually some monomial $x^np^m$. This constraint is equivalent to setting \eqref{eq:ehrenfest} to 0, at least for operators without explicit time dependence. We are saying that the expectation value of these operators is time-independent, which constrains us to eigenstates--or, more generally, time-independent density matrices. This is just the statement that in eigenstates, time evolution is just a pure phase rotation.

Given the discussion in the previous sections, the expression in \eqref{eq:ehrenfest} should ring some alarm bells. Specifically, the quantity $\pex{[H,A]}$ is only well-defined if $\psi \in D(H) \cap D(A)$. This is a very strong assumption! Without this assumption, one must be more careful. Let us assume $\psi$ is an eigenstate of $H$, so $\psi \in D_a(H)$ and $H\psi = E\psi$. Then, due to the eigenvalue equation, the following expression vanishes so long as $\psi \in D(A)$:
\begin{equation*}
    (H\psi,A\psi) - (\psi,AH\psi) = E(\psi,A\psi) - E(\psi,A\psi) = 0
\end{equation*}
However, this is \textit{not} equivalent to the quantity $(\psi,[H,A]\psi)$. The commutator $[H,A]$ algebraically generates a new operator. There is no guarantee that $\psi$ is in the domain of this new operator. The correct relation is instead
\begin{equation}\label{eq:anom1}
    (H\psi,A\psi)- (\psi,AH\psi) = 0 = (\psi,[H,A]\psi) + \pex{(H^\dagger - H)A}
\end{equation}
The first term is the algebraic commutator extended to $D(H)$. But there is now an extra term $\mathcal{A} \equiv \pex{(H^\dagger - H)A}$. This modification to the Ehrenfest theorem has been noticed before in the literature \cite{anomaly, anomaly1}. It is dubbed an `anomaly', which is an appropriate term for a number of reasons. First, like the chiral anomaly in gauge theory, it is a total derivative term. It also appears as an additive modification to the `normal' Ehrenfest theorem \eqref{eq:ehrenfest}; one can consider this a 0+1-dimensional anomalous Ward identity. Dealing with operator domains is genuinely a quantum effect that alters conservation equations---an anomaly. 

Note that $\mathcal{A} = 0$ when $A$ keeps $D(H)$ invariant, as $H^\dagger = H$ on $D(H)$ by construction. But this is often not the case. By using the constraint $\pex{[H,\OO]} = 0$, we were unwittingly extending the algebraic commutator to the whole space $D(H)$. For the bootstrap program, the correct constraint to use is \eqref{eq:anom1}. One should evaluate the commutator $[H,A]$ algebraically and evaluate the anomaly term $\mathcal{A}$ for states $\psi \in D(H)$. 

\subsection{A correct recursion}
Let us apply these new ideas to generate a complete recursion for bootstrapping the positional moments on the half line. Including the anomaly, the bootstrap recursion is generated by the following constraint on operators $\OO$ in energy eigenstates $\psi \in D_a(H)$, which we take to be real: 
\begin{equation}\label{eq:newconstraint}
    0 = (\psi,[H,\OO]\psi) + \pex{(H^\dagger - H)\OO}
\end{equation}
Let us take the Hamiltonian to be $H = p^2 + V(x)$ and our first trial operator as $\OO_1 = x^n$. The algebraic commutator is
\begin{equation*}
    [H,\OO_1] = [p^2,x^n] = -2inx^{n-1}p - n(n-1)x^{n-2}
\end{equation*}
where we will always `normal order' $x$ in front of $p$. We can evaluate the anomaly by explicitly integrating by parts: 
\begin{align*}
    \mathcal{A}_1 &= \pex{(H^\dagger - H)x^n} = \intplus -\psi''x^n \psi + \psi\partial_x^2(x^n\psi)\\
    &= -\intplus \psi''x^n \psi + \left[\psi\partial_x(x^n\psi) - \psi'x^n\psi\right|^\infty_0 + \intplus \psi'' x^n \psi\\
    &= -\lim_{x\to 0}nx^{n-1}\psi^2\\
    \mathcal{A}_1 &= -\delta_{n,1}\psi(0)^2
\end{align*}
As promised, the anomaly is a surface term, picking up a dependence on the state boundary conditions. The result is a modified constraint which will help build the recursion:
\begin{equation}\label{eq:constr1}
    0 = 2in \pex{x^{n-1}p} + n(n-1)\pex{x^{n-2}} + \delta_{n,1}\psi(0)^2
\end{equation}
We can proceed the same way using the trial operator $\OO_2 = x^np$. The algebraic commutator is
\begin{equation*}
    [H,\OO_2] = -2inx^{n-1}p^2 - n(n-1)x^{n-2}p + ix^nV'(x)
\end{equation*}
while the anomaly term may be evaluated to yield
\begin{equation*}
    \mathcal{A}_2 = i\delta_{n,1}\psi(0)\psi'(0)
\end{equation*}
There is a special case when $n=0$. In that case, we need to evaluate
\begin{eqnarray}
\vev{(H^\dagger -H) p}&=&i \int_0^\infty  - \psi^{\prime\prime} \psi^\prime + \psi\psi^{\prime\prime\prime}= -i (\psi')^2 |_0^\infty +i  \psi \psi^{\prime\prime}|_0^\infty\nonumber\\
&=& i (\psi'(0))^2 +i \psi(0)^2 (E -V(0))   
\end{eqnarray}
where we used the Schr\"odinger equation to relate the second derivative to $\psi$. 

The result is another modified constraint:
\begin{equation}\label{eq:constr2}
    0 = 2in\pex{x^{n-1}p^2} + n(n-1)\pex{x^{n-2}p}- i\pex{x^nV'(x)} - i\delta_{n,1}\psi(0)\psi'(0)
\end{equation}
To generate the full recursion relation, we use \eqref{eq:constr1}, \eqref{eq:constr2} and the eigenvalue equation:
\begin{equation}
    \pex{x^{n-1}p^2} = E_\psi\pex{x^{n-1}} - \pex{x^{n-1}V(x)}
\end{equation}
The result is the following recursion relation for problems on the half line, with boundary conditions $\psi(0) + a\psi'(0) = 0$ in a given state for $n>0$: 
\begin{multline}\label{eq:realhalflinerec}
    0 = 2nE_\psi\pex{x^{n-1}} + \frac{1}{2}n(n-1)(n-2)\pex{x^{n-3}} \\
    - 2n\pex{x^{n-1}V} - \pex{x^nV'} + \delta_{n,2}\psi_0^2 + \delta_{n,1}\frac{\psi_0^2}{a}\qquad\qquad\qquad\qquad
\end{multline}
and
\begin{equation}
0 = -\vev{V'}_\psi + (\psi'_0)^2+\psi_0^2(E-V(0))
\end{equation}
for $n=0$.
In these  equations $\psi_0 \equiv \psi(0)$. By including the anomaly terms, the recursion is now sensitive to the choice of operator domain for the Hamiltonian. We see that the recursion \eqref{eq:airyrec} that we used for bootstrapping the linear potential omitted these contact terms, which amounted to setting $\psi_0 = 0$; a specific choice of operator domain. Also we chose $\vev 1=1$ without paying attention to it,  but without the anomaly term, we would have obtained the contradiction $\vev 1=0$. 
That is why the results uncovered only the Dirichlet energy spectrum! In the next section, we will revisit the Airy problem and apply these results to the numerical bootstrap. 

\subsubsection{Bound states and the delta function}
As usual, the $n = 1$ case of the recursion \eqref{eq:realhalflinerec} gives us the virial theorem: 
\begin{equation*}
    E_\psi = \pex{V} + \frac{1}{2}\pex{xV'(x)} + \frac{1}{2}\psi_0\psi'_0
\end{equation*}
There is now an anomalous contribution to the energy, one which vanishes in the case of either pure Dirichlet or Neumann boundary conditions. Interestingly, this contribution persists when $V = 0$. Let us consider this free particle on a half line. The recursion suggests there should be a state with energy
\begin{equation*}
    E_a = -\frac{1}{2a}\psi_0^2
\end{equation*}
This state is created by the boundary conditions. It is also exactly the same as the energy of a state bound in an inverted delta function potential on $\RR$ (see Appendix A, also \cite{anomaly}).  

This gives us some physical insight to the situation regarding the anomaly: the boundary conditions at the origin are like adding a delta function source. This delta function source must come with a dimensionful parameter, e.g. a scale, for the Hamiltonian to be dimensionally consistent. The free particle on the half line does not have translation invariance, but it does have dilatation, or scale, invariance. In the quantum theory, the boundary conditions introduce a dimensionful parameter, breaking the classical scale invariance of the system. This is thus the simplest possible example of a conformal anomaly. 

\subsection{The Airy Bootstrap (Slight Return)}
Let us consider the linear potential again, this time using the anomaly-corrected recursion \eqref{eq:realhalflinerec}. Our recursion is thus, for $n>0$,  
\begin{equation*}
    \ex{n} = \frac{1}{2n+1}\left[ 2nE\ex{n-1} + \frac{1}{2}n(n-1)(n-2)\ex{n-3} + \delta_{n,2}\psi_0^2 + \delta_{n,1}\frac{\psi_0^2}{a}\right]
\end{equation*}
In the case of Dirichlet boundary conditions $\psi_0 = a = 0$, both contact terms vanish and we are left with the recursion \eqref{eq:airyrec}. This gave us a one-dimensional search space $\{E\}$ which correctly yielded the Dirichlet spectrum. In the case of Neumann conditions $a \to \infty$, one of the contact terms persists while the other vanishes, and the recursion depends also on $\psi_0$. 

We can consider the case of Neumann BCs as a two-dimensional bootstrap search space $\{E,\psi_0\}$. By borrowing the methods of \cite{Berenstein:2021loy}  for the double well potential, we can perform the numerical bootstrap by searching for points which pass the positivity checks in the $(E,\psi_0)$ plane. This bootstrap should recover both the Dirichlet and Neumann spectra (and no others); allowed islands which form along the axis $\psi_0 = 0$ should do so at the Dirichlet energy levels.
\begin{figure}[h!]
    \centering
    \includegraphics[scale = 0.6]{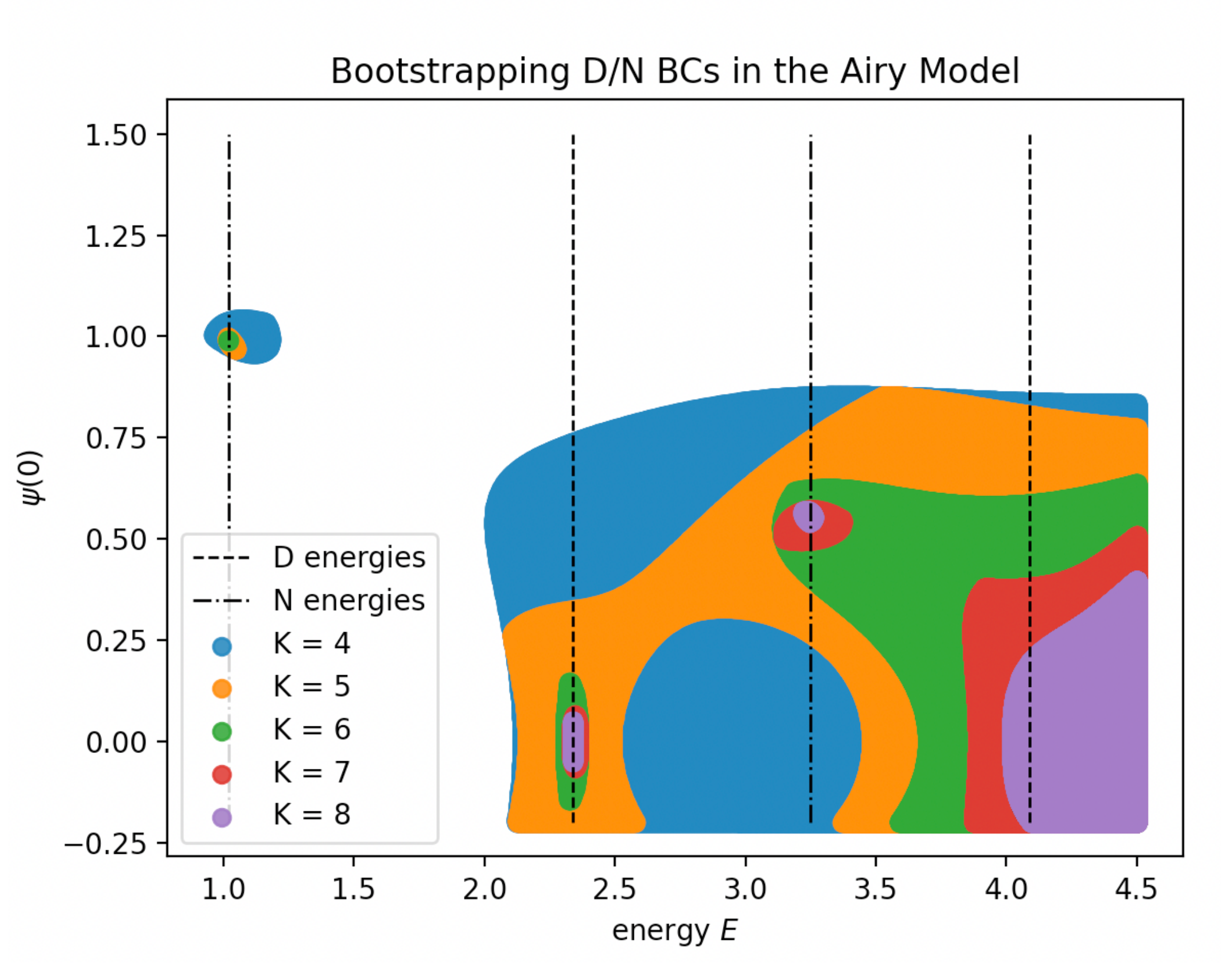}
    \caption{Bootstrapping the Dirichlet and Neumann spectra by passing to a search space of dimension two, with free parameters $E,\psi_0$. }
    \label{fig:airyDN}
\end{figure}

The result is shown in Fig. \ref{fig:airyDN}. The bootstrap correctly finds the Dirichlet levels and the Neumann levels, while not returning results for states with mixed boundary conditions: this is expected as we did not include both contact terms. To find the mixed spectra, we could increase the dimension of the search space once more, bootstrapping the free parameters $\{E,\psi_0,\psi_0'\}$ (bootstrapping $a$ would require $a \in (-\infty,\infty)$ which is computationally undesirable). In this way we can fully specify the desired boundary conditions for any problem on the half line via the recursion \eqref{eq:realhalflinerec}. We note that in the current release of \textit{Mathematica}, the native differential eigensystem solver can only handle homogeneous Dirichlet/Neumann boundary conditions. While the algorithmic implementation of the bootstrap here is much slower, it is already capable of solving a wider class of problems, especially if one is only interested in low-lying energies. 

\begin{figure}
    \centering
    \includegraphics[scale = 0.5]{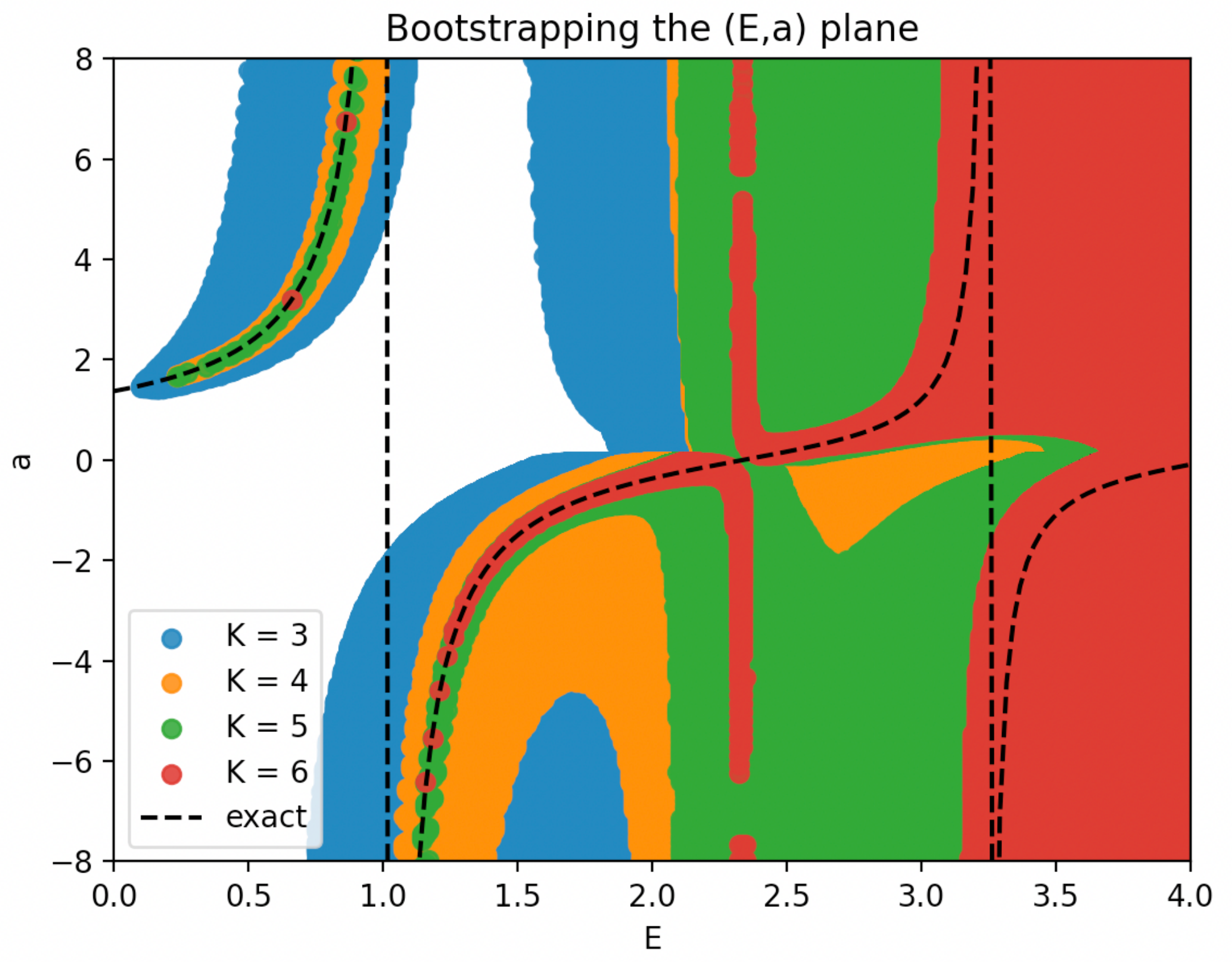}
    \caption{Projecting a 3d bootstrap into the $(E,a)$ plane. While numerically sparse, the bootstrap data agrees with the analytical predictions for the Robin boundary conditions. The vertical signal at $E \approx 2.33$ corresponds to the Dirichlet ground state. }
    \label{fig:airyplane}
\end{figure}

To demonstrate that the bootstrap can correctly find the full Robin boundary conditions, we can perform a low-resolution search for positivity in the 3d space of $\{E,\psi_0,\psi_0'\}$ then project down into the $\{E,a\}$ plane by taking $a = -\psi_0/\psi_0'$. We can analytically compute the dependence of the eigenvalue $E$ on the parameter $a$. The normalizable solution of $-f'' + xf = Ef$ is $f \propto \Ai(x-E)$. So we should have $a(E) = -\Ai(-E)/\Ai'(-E)$. The results are shown in Fig. \ref{fig:airyplane}. Due to the larger dimension of the search space, getting numerically satisfactory results is computationally intensive, at least done in the most naive, brute-force way. But the `experimental' data clearly conforms with our analytical expectations. This verifies that the anomalous contributions in the recursion do correctly account for the Robin boundary conditions.

\section{Conclusion}

The power of the bootstrap approach to quantum mechanics is due to its reliance on the algebraic structure of the problem. The benefit of this approach is that the algebraic structure is usually known \textit{a priori}, and no explicit knowledge of the dynamics is required.
 We rely on positivity constraints inherited from the unitarity of representations, and linear constraints are furnished by the commutation relations  between the operators in question.
 
We showed how sometimes we needed to supplement the na\"\i ve moment problem relations with additional physical constraints on moments of inverse powers of functions.
 These additional constraints were able to impose physically sensible boundary conditions at the origin: sufficient fast decay  at the origin, or vanishing of the measure on a small interval around 
 the origin. Correctly including these constraints allowed us to cut down areas of the search space which in previous work remained unconstrained. 
 Some of these constraints are mathematically required while others serve to disqualify unphysical solutions.

For problems on the half line, the standard treatment in the Schrodinger picture requires one to supply boundary conditions to solve the differential equation. 
It is not so obvious how one might interpret these data from an algebraic perspective on quantum mechanics. The correct interpretation is that 
of defining domains of self-adjointness for the Hamiltonian and its constituent operators.  By carefully considering these definitions, we were able to extract 
anomalous contributions to the bootstrap recursion which allowed us to fully specify any mixed linear boundary condition for a state on the half line. 

The fact that these anomalous contributions arose from demanding hermiticity of the Hamiltonian is quite fitting for the bootstrap story, where the fundamental constraint is unitarity. 
Unitarity gives us the positivity condition on the moment matrices. Unitarity of time evolution in turn requires that the Hamiltonian is self-adjoint. One can really regard these anomalous contributions to the bootstrap recursion as another 
constraint inherited from unitarity. 

Despite our ability to fully specify bootstrap data for half line problems, the situation needs further study for problems on an interval or the cirlce. 
For the problem on the circle, the techniques used here do not reveal any missing contributions, as we verify in Appendix \ref{sec:AppendixC}. Other 
methods \cite{Tchoumakov:2021mnh} are required in order to properly extract i.e. the quasimomentum dispersion relation for periodic potentials. 

While our numerical methods that have been used work well enough as a proof-of-concept, they are still mostly naive and scale very inefficiently with the dimension 
of the search space. The problem is algorithmic: how to find good regions of the search space before doing positivity tests.

One approach to mitigating this effect might be to try to translate the bootstrap problem into a `semi-definite program'. This is a well-studied class of convex optimization problems. 
With such an approach, the matrix of correlations, e.g. $M_{ij} = \ex{i+j}$, becomes the optimization variable. The objective function is a scalar defined over the cone of positive semi-definite matrices, linear in the matrix elements. 
Solvers for these types of problems employ various algorithms to make searching the high-dimensional space tractable. One may then try to add constraints to the optimization. 
Generally, such constraints should be linear in the correlation matrix elements. However, the bootstrap recursion for eigenstates \eqref{eq:realhalflinerec} is non-linear. 
This has been dealt with in the literature by relaxing the non-linear equality to an inequality on a new matrix variable \cite{Kazakov:2021lel, Kazakov:2022xuh}. 
SDP solvers have been applied to the quantum mechanical bootstrap to numerically bound the ground state energies from below \cite{Lawrence:2021msm}, both in one-particle systems and multi-site spin chains, however, they are not yet set up to deal with 
the full spectral problem. 
 
\acknowledgments
We would like to thank R. Brower, Y. Meurice, J.A. Rodriguez  for discussions and correspondence.
We would also like to thank K. Siampos, K. Sfetsos, G. Itsios for correspondence on the conventions of the implementation of our code.
Research supported in part by the Department of Energy under grant DE-SC0019139.

\appendix

\section{The delta-function potential}
Consider the Hamiltonian with a delta function potential:
\begin{equation}
    H = -\frac{d^2}{dx^2} - \frac{1}{a}\delta(x)
\end{equation}
We can just use the recursion \eqref{eq:oldrec} for moments of distributions on the real line. When we evaluate the expectation values of the potential, the delta-function will pick up residues of the state $\psi$.
\begin{align*}
m\langle x^{m-1}V(x)\rangle &= -\frac{m}{a}\int_\RR dx\ x^{m-1}\psi(x)^2\delta(x)\\
&= -\frac{1}{a} \lim_{x\to0}\ mx^{m-1}\psi(x)^2\\
\implies \quad m\langle x^{m-1}V(x)\rangle&= -\frac{1}{a} \delta_{m,1}\psi_0^2
\end{align*}
where $\psi_0 \equiv \psi(0)$ and we're now using the Kronecker delta. Similarly, 
\begin{align*}
    \langle x^m V'(x)\rangle &= -\frac{1}{a} \int_\RR dx\ x^m \frac{d}{dx}[\delta(x)]\psi^2\\
    &= \frac{1}{a} \int_\RR dx\ \delta(x)\frac{d}{dx}[x^m\psi^2]\\
    &= \frac{1}{a} \int_\RR dx\ \delta(x)[mx^{m-1}\psi^2 + 2x^m\psi\psi']\\
    \implies \quad \langle x^m V'(x)\rangle&= \frac{1}{a}\delta_{m,1}\psi_0^2 \\
\end{align*}
where we integrate by parts to make sense of the distributional derivative, and are considering only $m \geq 1$. This results in a recursion relation with a contact term:
\begin{equation}\label{eq:dfuncrec}
    0 = 2mE\ex{m-1} + \frac{1}{2}m(m-1)(m-2)\ex{m-3}  + \frac{1}{a} \delta_{m,1}\psi_0^2
\end{equation}
which furnishes constraints for $m \geq 1$. The known solution for an inverted delta-function potential is $\psi(x) \sim e^{-\kappa |x|}$. Hence the moments should grow approximately like gamma functions, and the derivative will be undefined at the origin. When $m=1$ the recursion gives
\begin{equation}\label{eq:dviri}
    E = -\frac{1}{2a}\psi_0^2
\end{equation}
This is the virial theorem. Continuing, the vanishing of the odd moments is guaranteed by the $m = 1$ case which sets $\ex{} = 0$. The rest of the even moments may be computed by a simple recursion for $n \geq 1$:
\begin{equation}
    \label{eq:dfuncshort}
    \ex{2n} = -\frac{n}{2E}(2n-1)\ex{2n-2}
\end{equation}
Note that positivity of the even moments requires $E <0$, which by \eqref{eq:dviri} requires $a > 0$. The bootstrap already tells us that normalizable states only live in the inverted delta potential.

We can actually solve this recursion explicitly. First, the normalization constraint fixes $\ex{2} = -(2E)^{-1}$. This then uniquely determines all higher moments $\ex{2n}$. The result is
\begin{equation*}
    \ex{2n} = (-1)^n\frac{(2n)!}{(4E)^n}
\end{equation*}
We know the wavefunction PDF to be even by symmetry. Consider the Fourier transform of the wavefunction PDF. It may be expressed as a power series in the moments:
\begin{equation*}
    \mathcal{F}[|\psi(x)|^2](k) = \int_{-\infty}^\infty dx\ e^{-ikx}|\psi|^2 = \sum_{m=0}^\infty  \frac{(-ik)^m}{m!}\ex{m} = \sum_{n=0}^\infty(-1)^n \frac{k^{2n}}{(2n)!}\ex{2n}
\end{equation*}
where we have used the vanishing of the odd moments in the last step. Using our expression for the even moments we can evaluate the sum as
\begin{equation*}
    \mathcal{F}[|\psi(x)|^2](k) = \sum_{n=0}^\infty \frac{k^{2n}}{(4E)^n} = \frac{4E}{4E - k^2}
\end{equation*}
Finally we invert the Fourier transform to obtain an explicit form of the wavefunction (PDF):
\begin{equation*}
    |\psi(x)|^2 = \sqrt{-E}e^{-2\sqrt{-E}|x|}
\end{equation*}
This is an example where the moment recursion solves the system explicitly.

\section{The bootstrap is well-behaved on $\RR$}\label{sec:apb}
In this section we quickly review some theorems which classify a large set of familiar, real-line Hamiltonians as self-adjoint, in the formal sense of operator domains discussed in section 4. Let us consider the following Hamiltonian on $L^2(\RR)$:
\begin{equation}\label{eq:rham}
    H = p^2 + V(x) = -\frac{d^2}{dx^2} + V(x)
\end{equation}
When the potential is real $V:\RR \to \RR$, a theorem of von Neumann \cite{reed2} states that the deficiency indices of the Hamiltonian are equal: $n_+ = n_-$. This means that such Hamiltonians are either essentially self-adjoint or admit self-adjoint extensions. Note that for states in $L^2(\RR)$, the Hamiltonian \eqref{eq:rham} is symmetric by virtue of the boundary conditions. 

Consider the space $C^\infty_c(\RR) \subset L^2(\RR)$ of smooth functions with compact support on $\RR$. If we let $D(H) = C^\infty_c(\RR)$, a theorem of Kato and Rellich \cite{hall,reed2} classifies the self-adjointness of $H$ based on properties of the potential $V(x)$. A special case of the theorem (see 9.39 in Hall \cite{hall}) is as follows. 

\textbf{Theorem.} Let $H = -\partial_x^2 + V(x)$ on the domain $D(H) = C^\infty_c(\RR)$. The operator $H$ is self-adjoint on $D(H)$ if the potential may be decomposed as $V = V_1 + V_2 + V_3$, where $V_1 \in L^2(\RR)$, $V_2$ is bounded, and $V_3 \geq 0$ is locally $L^2$. $\Box$

This class of potentials includes any potential which is smooth and $V(x) \to \infty$ as $|x| \to \infty$. This would include the harmonic potential, the double well potential, etc. Any ``confining" potential in which the classical physics is bounded should lead to a self-adjoint Hamiltonian. Thus, no extra work is needed. The questions of boundary conditions are answered by the requirement of compact support. Bootstrapping such problems consists of just checking the Hamburger matrix. 

\section{Self-adjoint domains on the interval}\label{sec:AppendixC}
Given that the inclusion of anomaly terms allowed us to specify boundary conditions for the half line bootstrap, one is tempted to ask if the same can be said for bootstrapping problems on the interval. Our approach \cite{ Berenstein:2021loy} did not allow us to specify the quasimomentum, so we could only detect energy bands. Others \cite{Aikawa:2021eai, Tchoumakov:2021mnh} came to the same conclusion, and tried new methods to obtain the full dispersion relation. 

For completeness, we will analyze the problem on an interval using the same approach as in the previous section: by precisely defining operator domains and analyzing the presence of possible anomaly terms. We find that unlike the half line, the recursion is insensitive to a large family of inequivalent boundary conditions on the interval. These families of boundary conditions have been studied in multiple contexts \cite{reed2,anomaly1}, and essentially correspond to the theory of Floquet exponents, or, to a condensed matter theorist, the Bloch quasimomentum. 

\subsection{Operator domains}
In the following, we will work over the Hilbert space $\HH = L^2[0,1]$ with the additional assumption that the states $\psi$ are smooth. However, we will not assume the states are real. Consider first the momentum operator $p = -i\partial_x$. It is symmetric when
\begin{equation*}
    (\psi,p\phi) - (p\psi,\phi) = 0 = \Bar{\psi_1}\phi_1 - \Bar{\psi}_0\phi_0
\end{equation*}
where we use e.g. $\phi_x = \phi(x)$. One choice of a symmetric domain is Dirichlet boundary conditions $\psi_1 = \psi_0 = 0$. However, it is not hard to see that in the case of Dirichlet boundary conditions, $D(p^\dagger)$ will not be constrained by any boundary conditions, and hence $D(p) \subset D(p^\dagger)$; $p$ will fail to be self-adjoint. However, symmetricity of $p$ is also achieved when $\phi_1 = e^{i\theta}\phi_0$ for all states in the domain $D(p)$. These are `twisted' boundary conditions, and include the periodic and anti-periodic sectors. Furthermore, it can be shown that $p$ is self adjoint on this domain \cite{anomaly1}. Hence, there is a family of suitable self-adjoint domains for $p$
\begin{equation}
    D_\theta(p) = \left\{\psi \in \HH \text{ smooth}\ |\ \psi_1 = e^{i\theta}\psi_0,\ \psi_0 \neq 0\right\} 
\end{equation}
The situation is somewhat similar for the Hamiltonian $H = p^2 + V = -\partial_x^2 + V$. The condition for symmetricity of $H$ is 
\begin{equation*}
    (H\phi,\psi) - (\phi,H\psi) = 0 = \left(\Bar{\phi}\psi')-\Bar{\phi}'\psi\right|^1_0
\end{equation*}
This is satisfied by Dirichlet boundary conditions. It is also satisfied if for $\forall\psi \in D(H)$ we have $\psi,\psi' \in D_\theta(p)$. However, in contrast to the momentum operator, both choices here will furnish a self-adjoint domain $D(H) = D(H^\dagger)$, as one can check by simply taking $\psi\in D(H)$ and $\phi \in D(H^\dagger)$ in the above. In conclusion, there is another one parameter family of self-adjoint domains for $H$: 
\begin{equation}
    D_\theta(H) = \left\{\psi \in \HH \text{ smooth}\ |\ \psi_1 = e^{i\theta}\psi_0,\ \psi_1' = e^{i\theta}\psi_0'\right\} 
\end{equation}
where the two ends of the interval are related to each other.
The other families amount to having general Robin boundary conditions at each end.

The physical difference between the twisted and Dirichlet boundary conditions is that of periodic potentials and the infinite square well. Let us focus on the former, and assume that we are imposing the twisted condition on our states. Note that with the Bloch ansatz $\psi_k(x) = e^{ikx}f(x)$, where $f$ is periodic, the state $\psi_k$ and all its derivatives are in the twisted sector of $D_\theta(H)$. 
\subsection{Anomalies?}
Now that we have defined our operator domains, it is natural to investigate whether there are anomaly terms that will modify the bootstrap recursion. We will proceed as in \cite{Berenstein:2021loy} to derive a recursion relation for the Fourier modes $t_n \equiv \pex{e^{2\pi i nx}}$. Recall that the anomaly-corrected constraint we use is 
\begin{equation}
    0 = (\psi,[H,\OO]\psi) + \pex{(H^\dagger - H)\OO}
\end{equation}
We will consider the twisted sector $D_\theta(H)$. The anomaly term vanishes whenever $\OO D_\theta(H) \subseteq D_\theta(H)$. Consider the operator $A_n = e^{2\pi inx}$. We can verify that for $\psi \in D_\theta(H)$, the state $\phi = A_n\psi \in D_\theta(H)$ also: 
\begin{equation*}
    \phi_1 = e^{2\pi in}\psi_1 = e^{i\theta}\psi_0 = e^{i\theta}\phi_0
\end{equation*}
Consider also the momentum operator $p = -i\partial_x$. Letting $\phi = p\psi$ for $\psi \in D_\theta(H)$, we have
\begin{equation*}
    \phi_1 = -i\psi'_1 = -ie^{i\theta}\psi'_0 = e^{i\theta}\phi_0
\end{equation*}
so that $\phi \in D_\theta(H)$ as well. As a result, all the operators $A_n,p,A_np$ leave the domain invariant, and hence do not contribute anomalies. These are the operators needed to create a recursion for the moments $t_n = \pex{e^{2\pi inx}}$, which is what we based our previous analysis on. For these choices of operators, we did not omit any anomaly contributions.

\bibliography{refs.bib}

\end{document}